\documentclass[10pt,twocolumn,letterpaper]{article}

\usepackage[pagenumbers]{cvpr} 

\usepackage{graphicx}
\usepackage{amsmath}
\usepackage{amssymb}
\usepackage{booktabs}
\usepackage{xcolor}

\usepackage[pagebackref,breaklinks,colorlinks]{hyperref}

\usepackage{pifont}
\usepackage{multirow}
\usepackage{makecell}

\newcommand{\bs}[1]{\boldsymbol{#1}}
\newcommand{\bh}[1]{\boldsymbol{\hat{#1}}}
\newcommand{\bm}[1]{\boldsymbol{\mathcal{#1}}}

\usepackage[capitalize]{cleveref}
\crefname{section}{Sec.}{Secs.}
\Crefname{section}{Section}{Sections}
\Crefname{table}{Table}{Tables}
\crefname{table}{Tab.}{Tabs.}

\def\cvprPaperID{8318} 
\def\confName{CVPR}
\def\confYear{2023}

\begin{document}

\title{Corner-to-Center Long-range Context Model for Efficient Learned Image Compression}

\author{Yang Sui\textsuperscript{1}\thanks{The work was done during the internship at Tencent America.} \enspace Ding Ding\textsuperscript{2}\enspace Xiang Pan\textsuperscript{2}\enspace Xiaozhong Xu\textsuperscript{2}\enspace Shan Liu\textsuperscript{2}\enspace Bo Yuan\textsuperscript{1}\enspace Zhenzhong Chen\textsuperscript{3}\thanks{The work was done during the visit at Tencent.} \\
\textsuperscript{1}Rutgers University \quad \textsuperscript{2}Tencent America \quad \textsuperscript{3}Wuhan University }

\maketitle


\begin{abstract}
    In the framework of learned image compression, the context model plays a pivotal role in capturing the dependencies among latent representations. To reduce the decoding time resulting from the serial autoregressive context model, the parallel context model has been proposed as an alternative that necessitates only two passes during the decoding phase, thus facilitating efficient image compression in real-world scenarios. However, performance degradation occurs due to its incomplete casual context. To tackle this issue, we conduct an in-depth analysis of the performance degradation observed in existing parallel context models, focusing on two aspects: the Quantity and Quality of information utilized for context prediction and decoding. Based on such analysis, we propose the \textbf{Corner-to-Center transformer-based Context Model (C$^3$M)} designed to enhance context and latent predictions and improve rate-distortion performance. Specifically, we leverage the logarithmic-based prediction order to predict more context features from corner to center progressively. In addition, to enlarge the receptive field in the analysis and synthesis transformation, we use the Long-range Crossing Attention Module (LCAM) in the encoder/decoder to capture the long-range semantic information by assigning the different window shapes in different channels. Extensive experimental evaluations show that the proposed method is effective and outperforms the state-of-the-art parallel methods. Finally, according to the subjective analysis, we suggest that improving the detailed representation in transformer-based image compression is a promising direction to be explored. 
\end{abstract}

\vspace{-4mm}
\section{Introduction}

\label{sec:intro}

\begin{figure}[ht]
    \centering
    \includegraphics[width=0.8\linewidth]{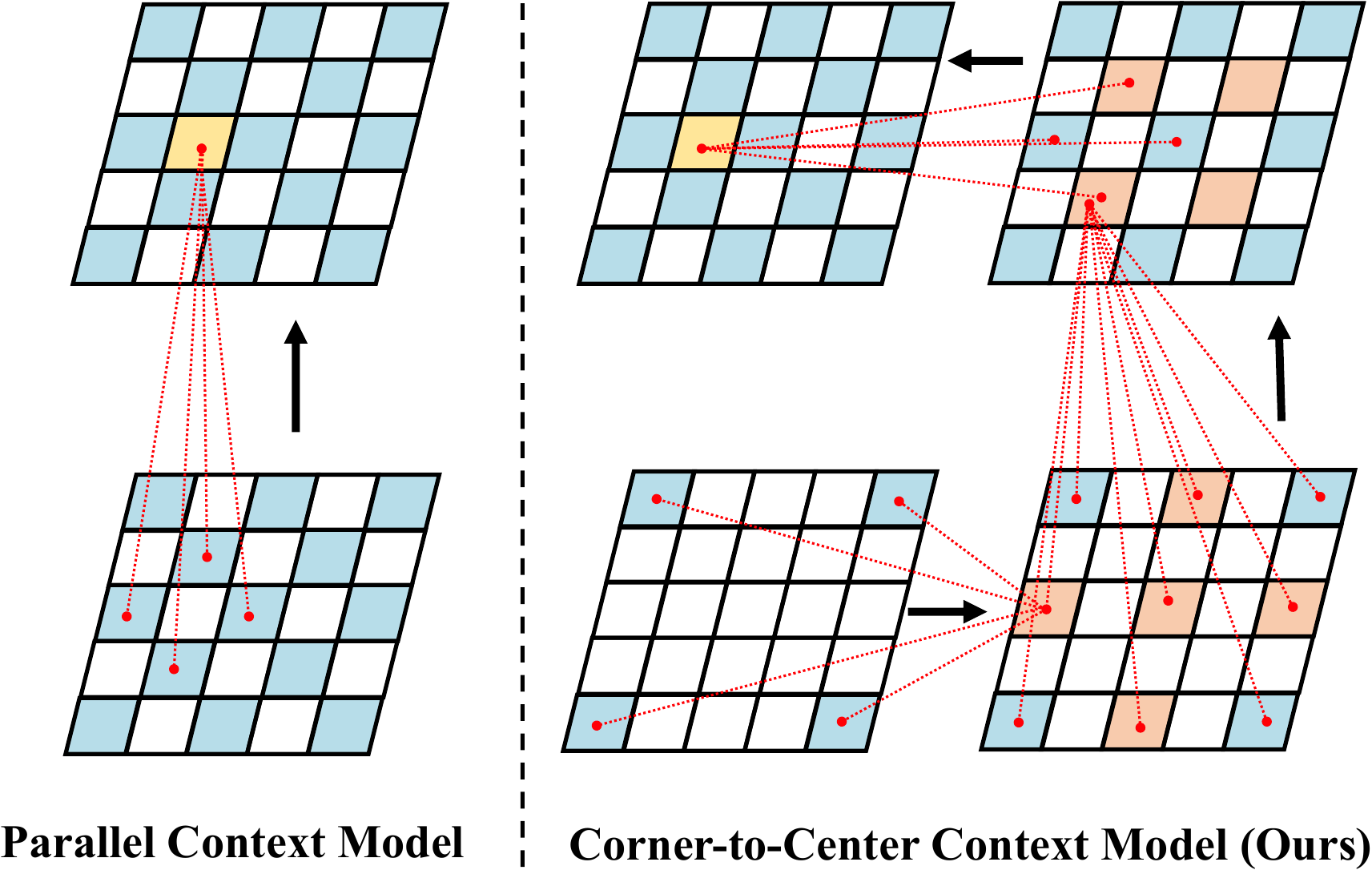}
    \caption{Different mechanisms between the parallel context model \cite{he2021checkerboard} and our Corner-to-Center Context Model (C$^3$M). Two different main points: (1) Different with parallel context model \cite{he2021checkerboard,qian2021entroformer} that almost 50\% latents (left blue grids, if large enough) are decoded by information from hyper-latents $\bh{z}$ without any causal context, our method has only a few corner points (right blue grids) that are decoded solely by hyper-latents $\bh{z}$. (2) Unlike parallel context model \cite{he2021checkerboard} that only examines the local limited receptive field when decoding the current $\hat{y}_{i}$, our method focuses on the long-range global area, which covers the whole latent features more effectively.}
    \label{fig:parallel-c3m}
    \vspace{-8mm}
\end{figure}

Image compression, a crucial and fundamental area of research in computer vision and signal processing domains, aims to represent the image data in a compact format while maximally preserving the information content. Recently, motivated by the remarkable representation capabilities of deep neural networks (DNNs), end-to-end learning-based image compression \cite{balle2017endtoend, balle2018variational, minnen2018joint} have demonstrated attractive rate-distortion (R-D) performance compared to the classical image codecs such as JPEG\cite{wallace1991jpeg}, JPEG2000\cite{taubman2002jpeg2000}, BPG\cite{bellard2016bpg}, and VVC \cite{bross2021overview}. The basic design philosophy of learning-based image compression involves constructing an auto-encoder framework \cite{hinton2006reducing} with the entropy minimization constraints \cite{theis2017lossy, balle2017endtoend, balle2018variational, minnen2018joint, lee2018context, sun2021interpolation, qian2021entroformer, golts2021image} to achieve the best R-D performance. Specifically, at the encoding stage, the advanced VAE-based framework utilizes a DNN-based transform as the main encoder to project the images to a low-dimensional latent space. Following quantization, the entropy estimation model predicts the distributions of latents, which are subsequently compressed into a bit stream using an arithmetic encoder aided by the estimated distribution. At the decoding end, the same entropy estimation model is applied to the arithmetic decoder to recover the latency information. This information is then fed into a DNN-based main decoder to reconstruct the original image.


Within the aforementioned learned image compression framework, a critical component is the autoregressive context model \cite{minnen2018joint}, which seeks to predict the unknown codes based on the availability of previously decoded ones. Owing to the context model's ability to effectively extract the important spatial correlation present in the latents, it can notably improve the rate-distortion performance of the learning-based image compression, such as the Peak-Signal-to-Noise Ratio (PSNR).

Despite the context model offering promising advantages in enhancing compression performance, it incurs significant deployment costs. In general, due to the consecutive prediction of latents, the corresponding computational complexity of the autoregressive model is of the order of $\mathcal{O}(n^2)$ ($n$ is the height or width of latents). Such high cost stems from the underlying scheme of the context model, which requires all previously decoded latents to predict a new latent. Consequently, the autoregressive context model is incapable of concurrent execution. To address this challenge, \cite{he2021checkerboard} proposes a two-pass checkerboard context calculation for parallel decoding. During the first pass, context parameters are employed to decode 50$\%$ uniformly distributed latents. Subsequently, In the second pass, the remaining 50$\%$ latents are decoded by a Masked CNN model that operates on neighboring local pixels. For each pass, all latents are decoded independently. Since there exists no relationship between concurrent decoding operations, the checkerboard context model can be executed with parallel processing to expedite execution speed.

Although this checkerboard-style processing mechanism demonstrates improved parallelism and has been incorporated into the state-of-the-art learning-based image compression \cite{qian2021entroformer}, it is still not the ideal solution for building a context model aimed at high-performance neural image codec. More specifically, during the first decoding pass, half latents are predicted without any information from their causal context, rendering them unable to capture the dependencies from the previously decoded context. During the second decoding pass, the model extracts image textures only within the local region, disregarding the global semantic information. Such incomplete causal context and restricted receptive field, by their nature, inherently limit the performance of the context model.

To address this challenge, in this paper, we propose an efficient and high-performance transformer-based context model, termed the \textit{Corner-to-Center Context Model} (C$^3$M). Specifically, we leverage the logarithmic-based prediction order to progressively predict the context feature, moving from corner to center positions. At each prediction, we incrementally increase the number of to-be-predicted context features, maximizing the total received information. In addition, as Transformers exhibit a superior ability to capture long-range dependencies \cite{dosovitskiy2020image, naseer2021intriguing}, we opt for a transformer-based design as the backbone architecture. Furthermore, to expand and enrich the receptive field during the analysis and synthesis transformation processes, we integrate the \textit{Long-range Crossing Attention Module} (LCAM) in the encoder/decoder to capture rich semantic information across diverse windows with distinct shapes. In summary, the contributions of this paper are as follows: 
\begin{itemize}
    \vspace{-2mm}
    \item We investigate the fundamental reasons for the performance degradation in the existing parallel context model from two perspectives: Quantity and Quality of the conditions.
    \vspace{-2mm}
    \item We propose a new transformer-based context model characterized by a corner-to-center processing mechanism, logarithmic-based prediction order, and gradual prediction of context features at each prediction operation. 
    \vspace{-2mm}
    \item To further enhance the representation capacity of analysis and synthesis transformation, we propose to assign distinct window shapes to separate group channels of the attention module, facilitating the capture of long-range and diverse semantic information.
    \vspace{-2mm}
\end{itemize}

\section{Related Work}
\label{sec:related}

\textbf{Learned Image Compression.} Recently, learning-based end-to-end image compression approaches have received considerable attention. Stemming from the pioneering work~\cite{balle2016end} that develops VAE-based image compression framework, \cite{balle2018variational} incorporates a hyperprior into the neural image compression architecture to effectively capture spatial dependencies in the latent representation, outperforming the classical JPEG \cite{marcellin2000overview} solution. Later, \cite{kim2022joint,lee2018context} advocates employing an autoregressive context model, a module that facilitates a better prediction of the mean and scale parameters of the distribution of latents. Built upon this foundational framework that consists of the main encoder/decoder, context model, and hyper-encoder/decoder, several variants have also been proposed through the application of diverse network topologies, training losses, and distribution modeling \cite{cheng2020learned, toderici2017full, johnston2018improved, cui2021asymmetric,choi2019variable,dosovitskiy2019you,sun2021interpolation, agustsson2019generative}

\textbf{Context Model.} Inspired by autoregressive generative models such as PixelCNN \cite{van2016conditional}, \cite{minnen2018joint,cheng2020learned,lee2018context} propose context models that sequentially predict the latent representations, leading to effective capture of spatial dependencies. More specifically, \cite{mentzer2018conditional} leverages the 3D-CNN to learn a conditional probability model of the latent distribution of the autoencoder, and \cite{liu2019non} proposes a 3-D mask convolution-based context model to capture the spatial and the cross-channel contexts. However, the context prediction of autoregressive models requires the complete decoding of previous causal contexts, resulting in limited parallelism. To address this issue, \cite{minnen2020channel} presents a channel-wise context model to minimize the serial processing in the context model. \cite{he2022elic} proposes the uneven spatial-channel contextual adaptive model to improve performance. \cite{he2021checkerboard} designs a computationally efficient context model to accelerate spatial-wise computation. Unfortunately, such a checkerboard-style processing scheme brings R-D performance degradation. \cite{qian2020learning} propose to incorporate the global reference into the context model, further increasing the computational burden.  \cite{qian2021entroformer,koyuncu2022contextformer} use the transformer as the backbone architecture to obtain a better probability estimation than the convolution-based context model. However, the lack of casual context limits the performance of the learning-based image compression.

\section{Preliminary}
\label{sec:prelim}

In general, the framework of deep learning-based image compression first maps the input image to a latent representation via the non-linear transform. Then the latents are encoded by an arithmetic encoder to generate the bit stream for efficient storage. To recover the original image, the bit stream is decoded by an arithmetic decoder and fed into the main DNN-based decoder to generate a reconstructed image. Specifically, suppose $g_a(\cdot)$, $g_s(\cdot)$ are the non-linear transforms. Let $\boldsymbol{x}$ and $\boldsymbol{\hat{x}}$ denote the original input and reconstructed images, respectively. Let $\boldsymbol{y}$ and $\boldsymbol{\hat{y}}$ denote the pre-quantized and quantized latent representation, respectively, then the deep learning-based image compression can be described as:
\vspace{-2mm}
\begin{equation}
\begin{aligned}
\boldsymbol{y} = g_a (\boldsymbol{x}), \bh{y} = Q(\boldsymbol{y}),\\
\bh{y} = \text{AD}(\text{AE}(\bh{y})), \bh{x} = g_s(\bh{y}),\\
\end{aligned}
\label{eqn:equ1}
\vspace{-2mm}
\end{equation}
where $Q(\cdot)$ is quantization operation, and $\text{AE}$ and $\text{AD}$ represent the arithmetic encoding and decoding processes, respectively. The reconstructed image $\boldsymbol{\hat{x}}$ is the output of the corresponding (inverse) transform. In addition, a hyper-prior is used as side information to estimate the mean and scale parameters of latents that are predicted from the entropy model. Suppose $\boldsymbol{z}$ and $\boldsymbol{\hat{z}}$ denote the pre-quantized and quantized hyper-prior, respectively, and then they can be obtained from a pair of hyper non-linear transforms (denoted as $h_a(\cdot)$ and $h_s(\cdot)$) with $\boldsymbol{y}$ as input. Furthermore, in order to boost the R-D performance, A context model is integrated into the hyper-prior framework. Specifically, given the context model as $g_{cm}(\cdot)$ and Gaussian parameters network as $g_{ep}$. The key computation to generate the $i$-th latent presentation is:
\vspace{-2mm}
\begin{equation}
\begin{aligned}
    \boldsymbol{z} = h_a (\boldsymbol{y}), \boldsymbol{\hat{z}} = Q(\boldsymbol{z}), \boldsymbol{\psi} = h_s(\bh{z}), \\
    \boldsymbol{\phi} = g_{cm}(\bh{y}_{<i}, \boldsymbol{\psi}), \boldsymbol{\mu}, \boldsymbol{\sigma} = g_{ep}(\boldsymbol{\psi},\boldsymbol{\phi}), \\    
\end{aligned}
\label{eqn:equ2}
\vspace{-2mm}
\end{equation}
where arithmetic encoding and decoding (AE and AD) is ignored for simplicity. Here $\boldsymbol{\psi}$ and $\boldsymbol{\phi}$ denote the context parameters and context features, as the input and output of the context model $g_{cm}(\cdot)$ with masked convolutions, respectively. The probability of latents can be modeled by a conditional Gaussian mixture model: 
\begin{equation}
\begin{aligned}
p_{\boldsymbol{\hat{y}}\mid \boldsymbol{\hat{z}}}(\hat{y}_i\mid \boldsymbol{\hat{z}}) &= \left( \mathcal{N}(\mu_i,\sigma_i^2 )*\mathcal{U}(-0.5, 0.5) \right) (\boldsymbol{\hat{y}}_i),\\
p_{\boldsymbol{\hat{y}}\mid \boldsymbol{\hat{z}}}(\boldsymbol{\hat{y}}\mid \boldsymbol{\hat{z}}) &= \prod_{i=1}^{} p_{\bh{y}\mid \bh{z}}(\hat{y}_{i} \mid \bh{z}),
\end{aligned}
\label{eqn:equ4}
\vspace{-2mm}
\end{equation}
where $\mu_{i}$ and $\sigma_{i}$ represent the mean and scale of Gaussian distribution of a single latent $\bh{y}_{i}$, respectively. Notice that $Q(\cdot)$ in Equ. \ref{eqn:equ1} and Equ. \ref{eqn:equ2} is approximated by adding uniform noise during training, making the loss differentiable. Finally, the training loss is calculated as the weighted sum of bit-rate and distortion: 
\vspace{-2mm}
\begin{equation}
\begin{aligned}
L &= R + \lambda \cdot D \\
&= \mathbb{E}_{\boldsymbol{x} \sim p_{\boldsymbol{x}}} \left[ -\log_{2} p_{\bh{y} \mid \bh{z}} (\bh{y} \mid \bh{z}) -\log_{2}{\bh{z}}\right] \\ 
&+ \lambda \cdot \mathbb{E}_{\boldsymbol{x} \sim p_{\boldsymbol{x}}}{\left[ d(\boldsymbol{x}, \boldsymbol{\hat{x}}) \right]}, \\ 
\end{aligned}
\label{eqn:equ3}
\vspace{-2mm}
\end{equation}
where bit-rate $R$ is the approximated compression rate of latent and hyper-latent representation, and the distortion $D$ denotes the quality gap between reconstructed images and original natural images using the target metric, e.g., mean squared error (MSE).


\section{Corner-to-Center Context Model}
\label{sec:method}

\begin{figure*}[ht]
    \centering
    \includegraphics[width=0.8\linewidth]{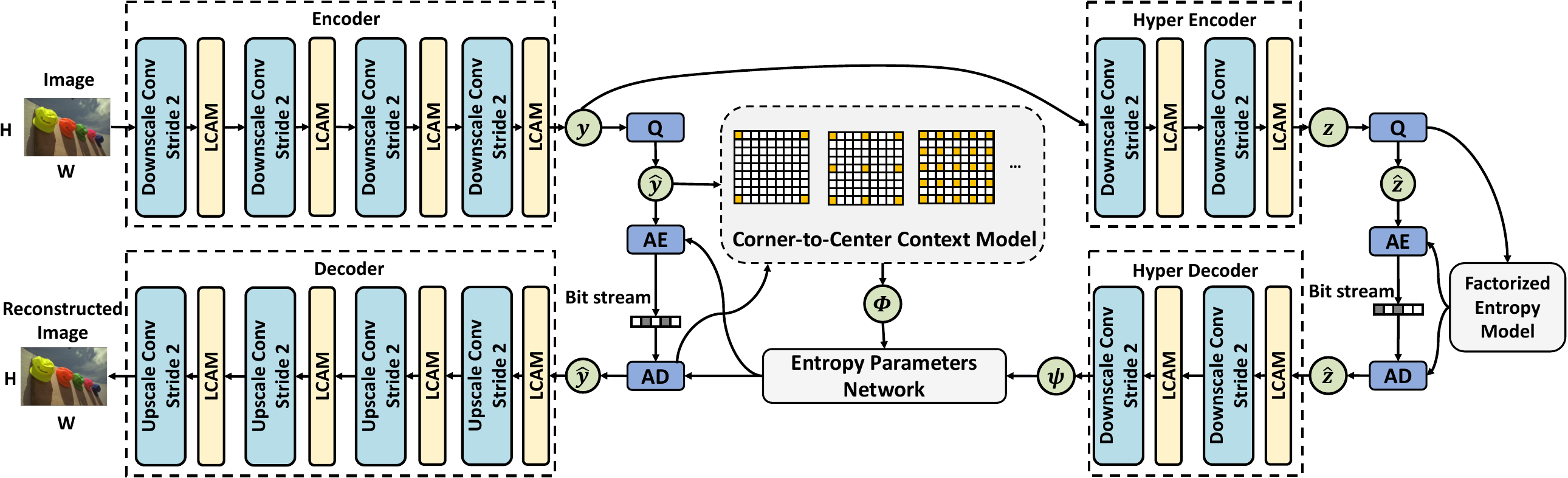}
    \caption{Our proposed learned image compression pipeline. Two advanced modules are proposed in this framework: (1) Corner-to-Center Context Model. (2) Long-range Crossing Attention Module (LCAM). }
    \label{fig:overview}
    \vspace{-6mm}
\end{figure*}


\subsection{Motivation}
\label{sec:motivation-context}
The existing \textbf{\textit{serial autoregressive model}} generates the Gaussian parameters of latent $\hat{y}_i$ based on the unknown causal context $\bh{y}_{<i}$. Given the context parameters $\boldsymbol{\psi}$ generated from hyper decoder $h_s(\cdot)$, it is first combined with context features $\boldsymbol{\phi}$ predicted by context model $g_{cm}(\cdot)$, and then fed into the entropy parameters network $g_\text{ep}(\cdot)$ to predict the mean and scale of Gaussian parameters of to-be-decoded latents $\boldsymbol{\hat{y}}$. However, such a serial processing scheme performs prediction in a pixel-by-pixel manner, requiring $\mathcal{O}(n^2)$ time complexity that is unacceptable when decoding a high-resolution image (e.g., about 2 minutes for decoding a 4K image). To reduce the decoding time, \cite{he2021checkerboard} proposes the \textbf{\textit{parallel context model}} to predict the mean and scale of latents $\hat{y}_i$ in two passes: 
\begin{equation}
\begin{aligned}
\mu_{i},\sigma_{i} = \begin{cases}
	g_{ep}(h_s(\boldsymbol{\hat{z}}), \boldsymbol{0})_{i}, & \text{first pass} \\
	g_{ep}(h_s(\boldsymbol{\hat{z}}), g_{cm}(\boldsymbol{\hat{y}}_{\text{half}(i)}; \boldsymbol{W}_{\text{conv}}))_{i}, & \text{second pass}
		              \end{cases}
\end{aligned}
\label{eqn:checker}
\end{equation}
where $\boldsymbol{\hat{y}}_{\text{half}(i)}$ denotes half uniformly distributed positions on the latents $\bh{y}$. $g_{cm}(\cdot)$ acts on the neighboring causal context of $\hat{y}_{i}$ within the receptive field of the convolutional layer in the context model, e.g., within 3$\times$3 kernel size. As shown in Eq. \ref{eqn:checker}, in the first pass, decoding 50$\%$ latents $\boldsymbol{\hat{y}}$ only leverages the information solely from context parameters $\bs{\psi} = h_s(\bh{z})$ without considering corresponding causal context $\boldsymbol{\hat{y}}_{<i}$ of each latent $\hat{y}_{i}$, denoted as $\boldsymbol{0}$. Without these causal contexts $\boldsymbol{\hat{y}}_{<i}$, it is unable to gather information from prior pixels, preventing the capture of the dependency between spatial correlation. Consequently, higher error and bias of predicted Gaussian parameters $\boldsymbol{\mu}, \boldsymbol{\sigma}$ from Entropy Parameters network $g_{ep}$ will be caused, further raising the bit-rate of encoded bit stream under the same image distortion level. For the second pass, when decoding the remaining 50$\%$ latents $\boldsymbol{\hat{y}}$, using CNN-based context model $\boldsymbol{W}_{\text{conv}}\in\mathbb{R}^{3\times3}$ only extracts the texture information in the local region. Compared to the autoregressive context model, the latents in the local receptive field of parallel context model, $\boldsymbol{\hat{y}}_{i} \in \mathbb{R}^{3\times3}$, is much less than $\boldsymbol{\hat{y}}_{<i}$ used in the autoregressive context model as shown in Fig. \ref{fig:parallel-c3m}. Such a small receptive field diminishes the capacity to capture semantic information from long-range dependencies. In addition, it also brings inaccuracy of the Gaussian parameters, causing the degradation of the rate-distortion performance.

To fully exploit the potential of the discarded information and capture the long-range dependencies based upon \textbf{\textit{parallel context model}}, two conditions should be satisfied:
    \vspace{-2mm}
\begin{itemize}
    \item \textit{Quantity Condition: A certain amount of previously decoded information is essential for estimating the unknown context.}
    \vspace{-2mm}
    \item \textit{Quality Condition: The decoded information should contain information that spans a broad range of the long-range casual context.} 
    \vspace{-2mm}
\end{itemize}
In other words, the \textit{Quantity Condition} requires more causal context $\boldsymbol{\hat{y}}_{<i}$ being fed into $g_{cm}$ instead of zero. The \textit{Quality Condition} requires the larger receptive field of context model $g_{cm}$. To satisfy the \textit{Quantity Condition} and \textit{Quality Condition}, we investigate the following two aspects: \ding{182} the \textbf{\textit{prediction strategy}}; \ding{183} the \textbf{\textit{backbone model}}. 



\begin{figure*}[ht]
    \centering
    \includegraphics[width=0.78\linewidth]{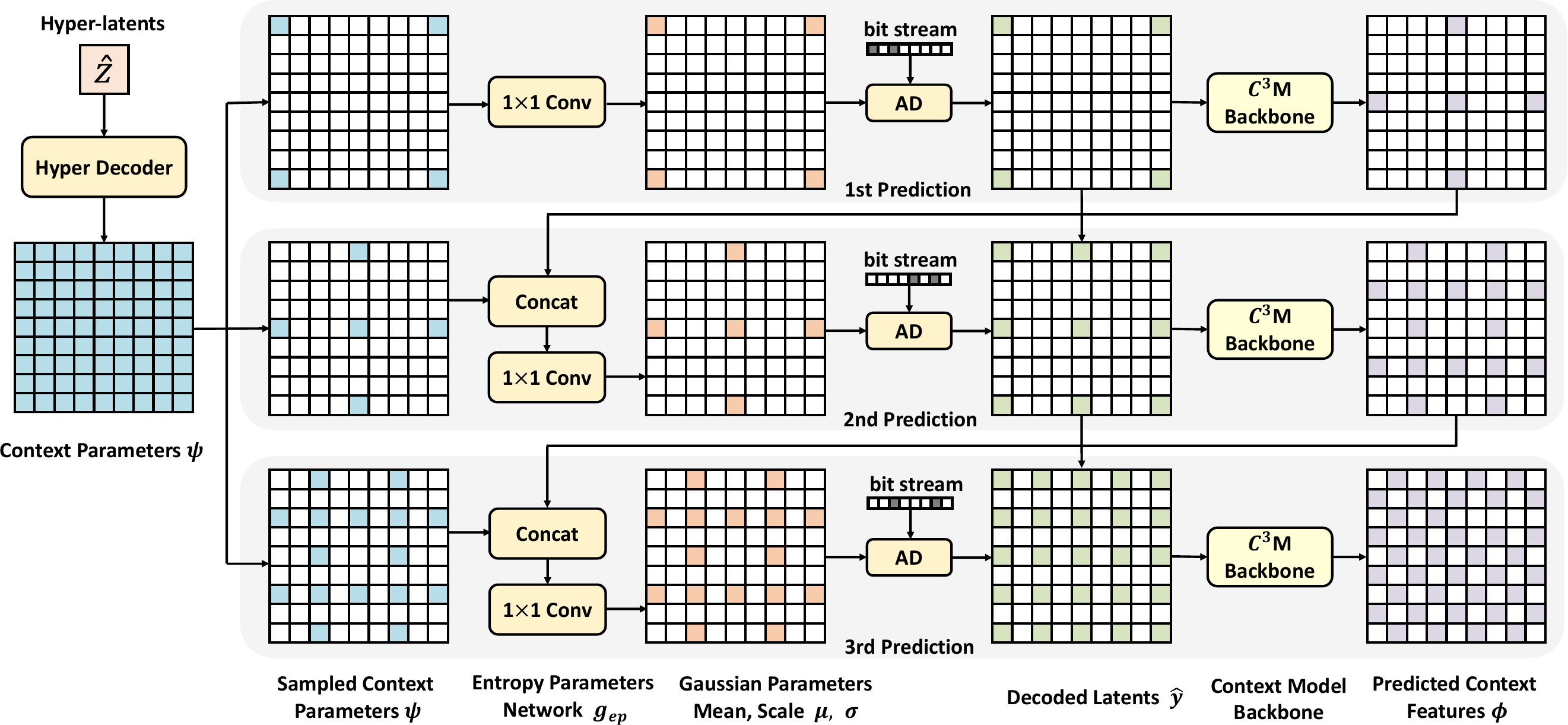}
    \caption{The prediction strategy of $\text{C}^3$M, including first three complete prediction steps. At each step, context parameters $\boldsymbol{\psi}$ are sampled and fed into the Entropy Parameters network $g_{ep}$ to predict the corresponding mean and scale. The latents $\bh{y}$ are decoded by sending the corresponding bit stream to an arithmetic decoder (AD) with the help of mean and scale. }
    \label{fig:c3m}
\vspace{-6mm}
\end{figure*}

\subsection{Proposed Solution}

\subsubsection{Prediction Strategy}

\textbf{\textit{Quantity Condition}.} For the Quantity Condition, 50$\%$ entropy parameters ($\boldsymbol{\mu}$ and $\boldsymbol{\sigma}$) in \cite{he2021checkerboard,qian2021entroformer} are solely predicted from context parameters $\boldsymbol{\psi}$. That means predicting each latent $\hat{y}_i$ in the first pass leverages \textbf{ZERO} causal context $\boldsymbol{\hat{y}}_{<i}$, failing to obtain the information of dependencies from previous decoded context, and thus it causes severe performance degradation. Evidently, a promising solution should make the majority of 50$\%$ entropy parameters take more causal context $\boldsymbol{\hat{y}}_{<i}$ into account, and only a few entropy parameters are derived purely from context parameters $\boldsymbol{\psi}$. 

Motivated by this philosophy, at each prediction step, we propose the progressively multi-stage prediction strategy as C$^3$M scheme: In the first step, we only predict less than $\tau\%$ ($\tau < 5$) entropy parameters $\boldsymbol{\mu}$, $\boldsymbol{\sigma}$ and referring latents $\boldsymbol{\hat{y}}$ without considering the causal context $\boldsymbol{\hat{y}}_{<i}$. In the second step, about $3\tau\%$ entropy parameters $\boldsymbol{\mu}$, $\boldsymbol{\sigma}$ and referring latents $\boldsymbol{\hat{y}}$ are predicted based upon their causal context $\boldsymbol{\hat{y}}_{<i}$ produced in the first step. Repeatedly, we gradually increase the prediction number of entropy parameters (e.g., from $3\tau\%$ to $100\%$) and their latents at each step until all entropy parameters and latents are constructed. 

\textbf{\textit{Quality Condition}.} To satisfy the Quality Condition, we should enlarge the receptive field of input and output of the context model at each prediction step. Since most semantic information lies in the center of the image instead of the edge, it is preferable that use less casual context to predict the edge pixels but more for center pixels. Therefore, at the early stages, the C$^3$M operates on the relative edge pixels of the image. As the decoding process proceeds, the effective area of the context model is gradually narrowed down from corner to center. Based on the logarithmic algorithm, we further extend and adapt it to the 2-D image compression tasks described in the following details. 


\textbf{Details.} As shown in Fig. \ref{fig:c3m}, given the hyper-latents $\hat{z}$, context parameters $\boldsymbol{\psi} \in \mathbb{R}^{h_{\psi} \times w_{\psi} \times c_{\psi}}$ are obtained through the hyper decoder $h_s(\cdot)$ as $\boldsymbol{\psi} = h_s(\boldsymbol{\hat{z}})$. For simplicity, here we denote $h_{\psi}, w_{\psi}$ as $h, w$. 
During the context modeling, at the first step, also known as the initialization step, the four-corner context parameters $\{\psi_{1,1}, \psi_{1,w},\psi_{h,1},\psi_{h,w}\}$ are sampled and fed into the Entropy Parameters network $g_{ep}$ to predict four pairs of corresponding mean and scale. Then, the latents $\{\hat{y}_{1,1}, \hat{y}_{1,w},\hat{y}_{h,1},\hat{y}_{h,w}\}$ are decoded by sending corresponding bit stream to an arithmetic decoder (AD) with the help of four pairs of mean and scale. Finally, it conveys the previously predicted latents $\{\hat{y}_{1,1}, \hat{y}_{1,w},\hat{y}_{h,1},\hat{y}_{h,w}\}$ into the our proposed corner-to-center context model. By using this prediction strategy, the model will predict the middle points among those latents that have already been decoded. Therefore, the context features $\{{\phi}_{1,\frac{w}{2}}, {\phi}_{1,\frac{h}{2}},{\phi}_{\frac{h}{2},\frac{w}{2}},{\phi}_{\frac{h}{2},w}, {\phi}_{h,\frac{w}{2}}\}$ are predicted based on the edge-latents $\{\hat{y}_{1,1}, \hat{y}_{1,w},\hat{y}_{h,1},\hat{y}_{h,w}\}$. 

In the second step, predicting the parts of remaining undecoded latents starts by sampling context parameters $\{{\psi}_{1,\frac{w}{2}}, {\psi}_{1,\frac{h}{2}},{\psi}_{\frac{h}{2},\frac{w}{2}},{\psi}_{\frac{h}{2},w}, {\psi}_{h,\frac{w}{2}}\}$ corresponding to the previously predicted context features. Then, we stack them with the channel dimension and send the combination into the $g_{ep}$ to predict five pairs of mean and scale corresponding to their positions. The latents $\{\hat{y}_{1,\frac{w}{2}}, \hat{y}_{1,\frac{h}{2}},\hat{y}_{\frac{h}{2},\frac{w}{2}},\hat{y}_{\frac{h}{2},w}, \hat{y}_{h,\frac{w}{2}}\}$ are decoded via sending the corresponding bit stream to an arithmetic decoder. Overall, we formulate the C$^3$M prediction strategy as follows: 
\begin{equation}
\begin{aligned}
\phi_{x^t,y^t} &= \begin{cases}
	g_{c^3m}(\hat{y}_{{x}^{t-1}_{1},{y}^{t-1}_{1}},\hat{y}_{{x}^{t-1}_{2},{y}^{t-1}_{2}},\ldots,\hat{y}_{{x}^{t-1}_{K},{y}^{t-1}_{K}} ), \\
	    \quad \quad \textit{if} ~ (x^t,y^t) \in \{ \left( \texttt{mid}(x_1, x_2),\texttt{mid}(y_1, y_2) \right) \mid \\ 
	    \quad \quad \quad \quad \quad \quad x_1, x_2 \in \boldsymbol{x}^{<t}, y_1, y_2 \in \boldsymbol{y}^{<t}\} 
	\\
	0, \quad \textit{otherwise}
              \end{cases} \\
\end{aligned}
\label{eqn:strategy1}
\end{equation}
where $\texttt{mid}(\cdot)$ denotes the operation to obtain the middle point. $\boldsymbol{x}^{<t}$ presents the $\boldsymbol{x}^{0}\cup\boldsymbol{x}^{1}\cup\ldots\cup\boldsymbol{x}^{t-1}$, where \{($\boldsymbol{x}^{t=0}, \boldsymbol{y}^{t=0})\} = \{ (1,1), (h,1),(1,w),(h,w)\}$. The corresponding mean and scale are calculated as:
\begin{equation}
\begin{aligned}
\mu_{x^t,y^t}, \sigma_{x^t,y^t} &= g_{ep}(\psi_{x^t,y^t},\phi_{x^t,y^t}) \\
\end{aligned}
\label{eqn:strategy2}
\end{equation}

Notice that the prediction phases are executed until all latents $\bh{y}$ are predicted. Here each generated ${\hat{y}_{i}}$ is mainly impacted by the nearest input latents. For instance, ${\hat{y}_{1,\frac{w}{2}}}$ is mainly impacted by ${\hat{y}_{1,1}}$ and ${\hat{y}_{1,w}}$ rather than ${\hat{y}_{h,1}}$ and ${\hat{y}_{h,w}}$. That follows the common concept that closer points own stronger relevance. 

\subsubsection{Backbone of C$^3$M: CNN or Transformer?}
\label{sec:contextbackbone}


The context models of the existing works can be divided into two main categories: CNN-based \cite{he2021checkerboard}, and Transformer-based \cite{qian2021entroformer}. The CNN-based context model is proposed to assist hyper-latents $\boldsymbol{\hat{z}}$ to predict the dependencies of adjacent pixels. Since the commonly used 3$\times$3 or 5$\times$5 convolutional kernels always focus on a small region near the to-be-predicted pixel, it can only capture the local dependencies with the texture information. On the other hand, as analyzed in \cite{graham2021levit, d2021convit}, transformer-based context models pay attention to the whole tokens of images, especially when predicting global dependencies with semantic information \cite{raghu2021vision, xiao2021early}. Considering 1) transformers integrate more global information than CNNs at lower layers and can preserve spatial information \cite{raghu2021vision}; and 2) transformers also outperform CNNs when exploiting the shape-based features, in this work we adopt a transformer-based context model to satisfy the Quality condition. 

\subsubsection{Details of C$^3$M Backbone}

Regarding the details of $\text{C}^3$M, we leverage the Masked Transformer model as our backbone. Given the reshaped input $\boldsymbol{\hat{y}} \in \mathbb{R}^{N \times d}$, where $N=H\times{W}$ denotes the total number of tokens sent into C$^3$M. Suppose the Masked Transformer model with the $\text{depth} = 1$, the context features $\boldsymbol{\phi}$ are predicted following:
\begin{equation}
\begin{aligned}
\boldsymbol{x^l} &= \text{Masked-MSA}(\text{LN}(\bh{y})) + \bh{y} \\
\boldsymbol{\phi} &= \text{MLP}(\text{LN}(\boldsymbol{x^l})) + \boldsymbol{x^l}
\end{aligned}
\label{eqn:transformer}
\end{equation}
where attention module in Masked-MSA is following \cite{qian2021entroformer}:
\begin{equation}
\begin{aligned}
\text{Attn}(\boldsymbol{Q},\boldsymbol{K},\boldsymbol{V}) = \text{softmax}(\frac{\boldsymbol{Q}\boldsymbol{K}^{\text{T}} + \boldsymbol{P}}{\sqrt{d}})\boldsymbol{V},
\end{aligned}
\label{eqn:trans}
\end{equation}
where $\boldsymbol{P}$ denotes the position encoding and we only calculate the $\boldsymbol{\phi}$. In our experiments, we stack the Masked Transformer model with $\text{depth} = 6$. It is worth noting that in the last stage of prediction, we adopt the 3$\times$3 CNN as the backbone context model with better capturing the nearby local texture information motivated by nature between CNNs and Transformer \cite{xiao2021early}.

\section{Long-Range Crossing Encoder/Decoder}
\label{sec:encdec}

In addition to the context model, the architecture of main and hyper codecs are also optimized in C$^3$M. Recall that the encoder $g_a(\cdot)$ and decoder $g_s(\cdot)$ aim to extract the low-dimension latents $\boldsymbol{y}$ of original image $\boldsymbol(x)$. To date, considering the transformer can better capture global information, the transformer-based encoder/decoder \cite{qian2021entroformer,zou2022devil} has been adopted in the state-of-the-art learning-based image compression. However, the existing solutions only focus on extracting global information while ignoring the local information, causing much loss of texture information. To that end, we propose a shape-fused local window as our attention scope to extract important global and local texture information simultaneously. Next, we describe our proposed approach for the main encoder as an example, and it can be generally applied to the main decoder and hyper encoder/decoder as well. 

\textbf{Encoder Pipeline.} 
To capture the texture information in the shallow layers and semantic information in the deep layers \cite{graham2021levit}, we fuse the convolutional and attention layer sequentially in one encoder block. As shown in Fig. \ref{fig:overview}. We first conduct the convolutional layer to downsample and tokenize an image $\boldsymbol{x}\in\mathbb{R}^{{H}\times{W}\times{C}}$ into $\boldsymbol{x}_{/2}\in\mathbb{R}^{\frac{H}{2}\times\frac{W}{2}\times{C}}$. The down-sample convolutional layer is set to 3$\times$3 kernel size, 128 channels with the stride of 2. Then, the tokens $\boldsymbol{x}_{/2}$ are sent to several Long-range Crossing Attention Module (LCAM), whose output $\boldsymbol{o}_{/2}\in\mathbb{R}^{\frac{H}{2}\times\frac{W}{2}\times{C}}$ will be the input of next block. Then, $\boldsymbol{o}_{/2}\in\mathbb{R}^{\frac{H}{2}\times\frac{W}{2}\times{C}}$ is fed into the several convolution layers with the stride of 2 and LCAMs in a repeated way. For deeper blocks, the number of channels of the convolutional layer and the LCAM increase from 128 to 192, and the depth from the first block to the fourth block are set to $2, 4, 6, 2$. Finally, we can obtain the low-dimensional latents $\bs{y} \in \mathbb{R}^{\frac{H}{16}\times\frac{W}{16}}\times{C}$ at the output end.

We provide a more detailed description of the Decoder Pipeline and Hyper Encoder/Decoder Pipeline utilized in our architecture in Appendix.
\begin{figure}[t]
    \centering
    \includegraphics[width=0.8\linewidth]{ 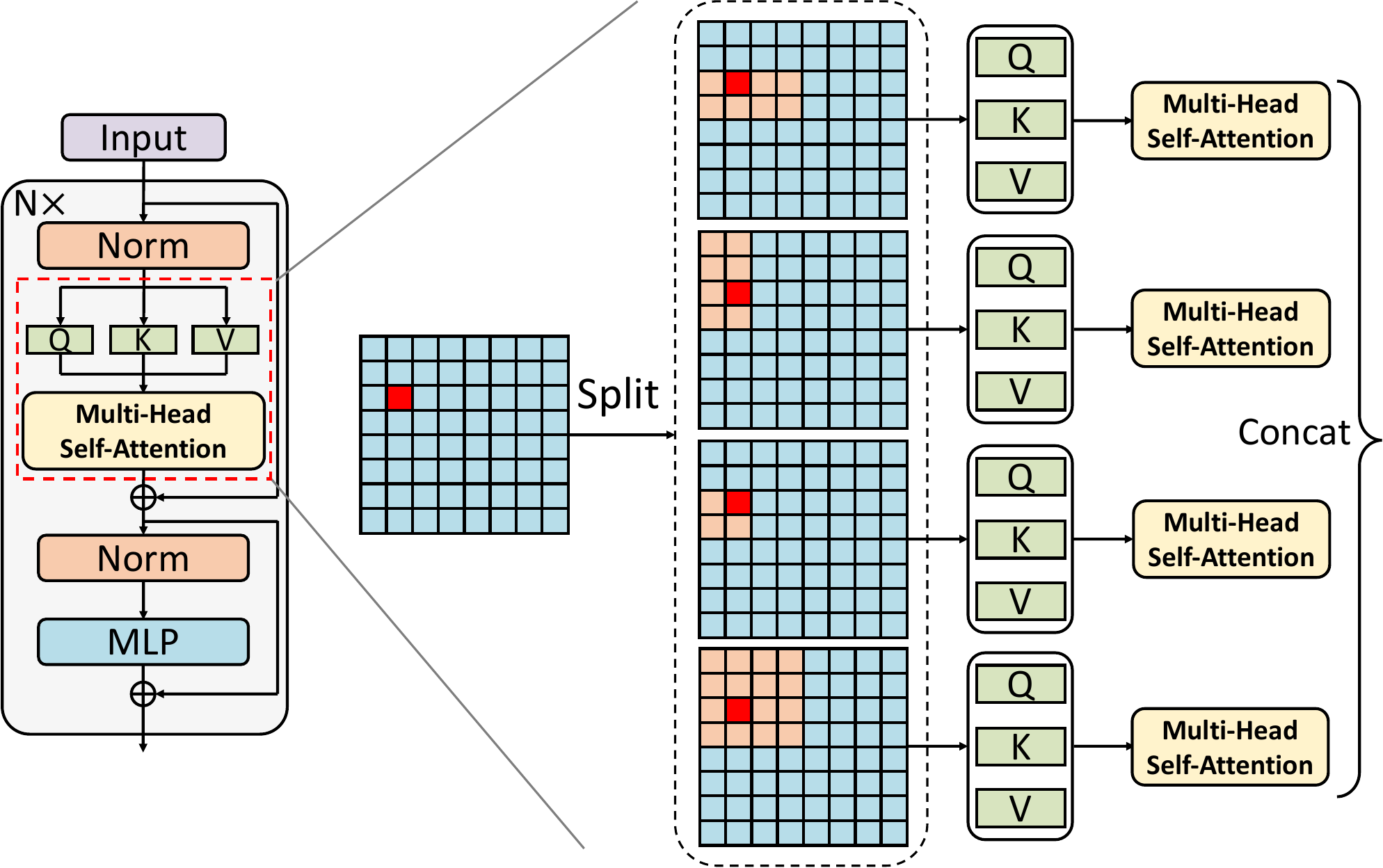}
    \caption{Long-range Crossing Attention Module.}
    \label{fig:lcam}
\vspace{-6mm}
\end{figure}

\textbf{Long-range Crossing Attention Module.} The major part of the long-range crossing attention module is identical to the regular transformer as described in Section \ref{sec:contextbackbone} except for the MSA module. As shown in Fig. \ref{fig:lcam}, for each long-range crossing attention module, we first reshape the input token $\bm{X}\in\mathbb{R}^{{H}\times{W}\times{C}}$ to $\bm{X}\in\mathbb{R}^{N\times{C}}$, where $N={H}\times{W}$. After performing the Layer Norm (LN), $\bm{X}$ is linearly projected by $\bm{W}_q, \bm{W}_k, \bm{W}_v$ to generate $\bm{Q}, \bm{K}, \bm{V}$. Then, we split the channels of $Q, K, V$ as:
\begin{equation}
\begin{aligned}
\bm{Q} &= [\bm{Q}^1,\bm{Q}^2,\bm{Q}^3,\bm{Q}^4] , \\
\bm{K} &= [\bm{K}^1,\bm{K}^2,\bm{Q}^3,\bm{K}^4] , \\
\bm{V} &= [\bm{V}^1,\bm{V}^2,\bm{Q}^3,\bm{V}^4] , \\
\end{aligned}
\label{eqn:split}
\vspace{-2mm}
\end{equation}
where each $\bm{Q}, \bm{K}, \bm{V}$ is divided to four groups in which $\bm{Q}^{i}$ has $C/4$ channels. Next, $\bm{Q}^{i}$ is evenly partitioned into the four shapes: horizontal ($i=1$), vertical ($i=2$), (K$\times$K) window($i=3$), (2K$\times$2K) window ($i=4$):
\begin{equation}
\begin{aligned}
\bm{Q}^{i} &= [\bm{Q}^{i}_{1},\bm{Q}^{i}_{2},\ldots,\bm{Q}^{i}_{M}] , \\
\end{aligned}
\label{eqn:split2}
\vspace{-2mm}
\end{equation}
where $\bm{Q}^{i}_{j}$ denotes the corresponding $j$-th specific window, and $\bm{K}^{i}$ and $\bm{V}^{i}$ are partitioned in the same manner. Then, a standard attention module is constructed as:
\begin{equation}
\begin{aligned}
\bm{Y}^{i}_{j} &= \text{Attn}(\bm{Q}^{i}_{j},\bm{K}^{i}_{j},\bm{V}^{i}_{j}), \\
\end{aligned}
\label{eqn:horizontal}
\vspace{-2mm}
\end{equation}
A concatenation scheme is then applied as follows:
\begin{equation}
\begin{aligned}
\bm{Y}^{i} &= \text{Concat}(\bm{Y}^{i}_{1},\bm{Y}^{i}_{2},\ldots, \bm{Y}^{i}_{M}), \\
\bm{Y} &= \text{Concat}(\bm{Y}^{1},\bm{Y}^{2},\ldots, \bm{Y}^{k}), \\
\end{aligned}
\label{eqn:concat}
\end{equation}
and finally, the projection is performed via a linear layer to calculate the output of LCAM as follows:
\vspace{-2mm}
\begin{equation}
\begin{aligned}
\text{LCAM}(\bm{X}) &= \bm{Y}\bm{W}^{0}, \\
\end{aligned}
\label{eqn:wo}
\vspace{-2mm}
\end{equation}
where $\text{LCAM}(\bm{X})$ is with the same shape of input $\bm{X}$.

\section{Experiments}
\label{sec:exp}

\textbf{Training Setup.} The proposed model is trained on Flicker2W \cite{liu2020unified} following the the strategy used in \cite{qian2020learning}. For the optimizer, we use Adam with the default hyper-parameters. The learning rate starts from 10$^{-4}$ and decreases to 10$^{-5}$ after 300 training epochs. The total number of training epochs is set to 400. We evaluate the proposed model by calculating the rate-distortion (RD) performance. For distortion terms, we adopt the MSE metric. For measuring the performance on the test dataset, the bit rate, in terms of average bits per pixel (BPP), and the distortion, in terms of PSNR and MS-SSIM, are evaluated as the compression quality metric. We train different models with different $\lambda$'s suggested in \cite{begaint2020compressai} to evaluate the MSE-oriented rate-distortion performance for various bit-rates. CompressAI is used as the framework \cite{begaint2020compressai}.

\begin{figure}[t]
    \centering
    \includegraphics[width=0.8\linewidth]{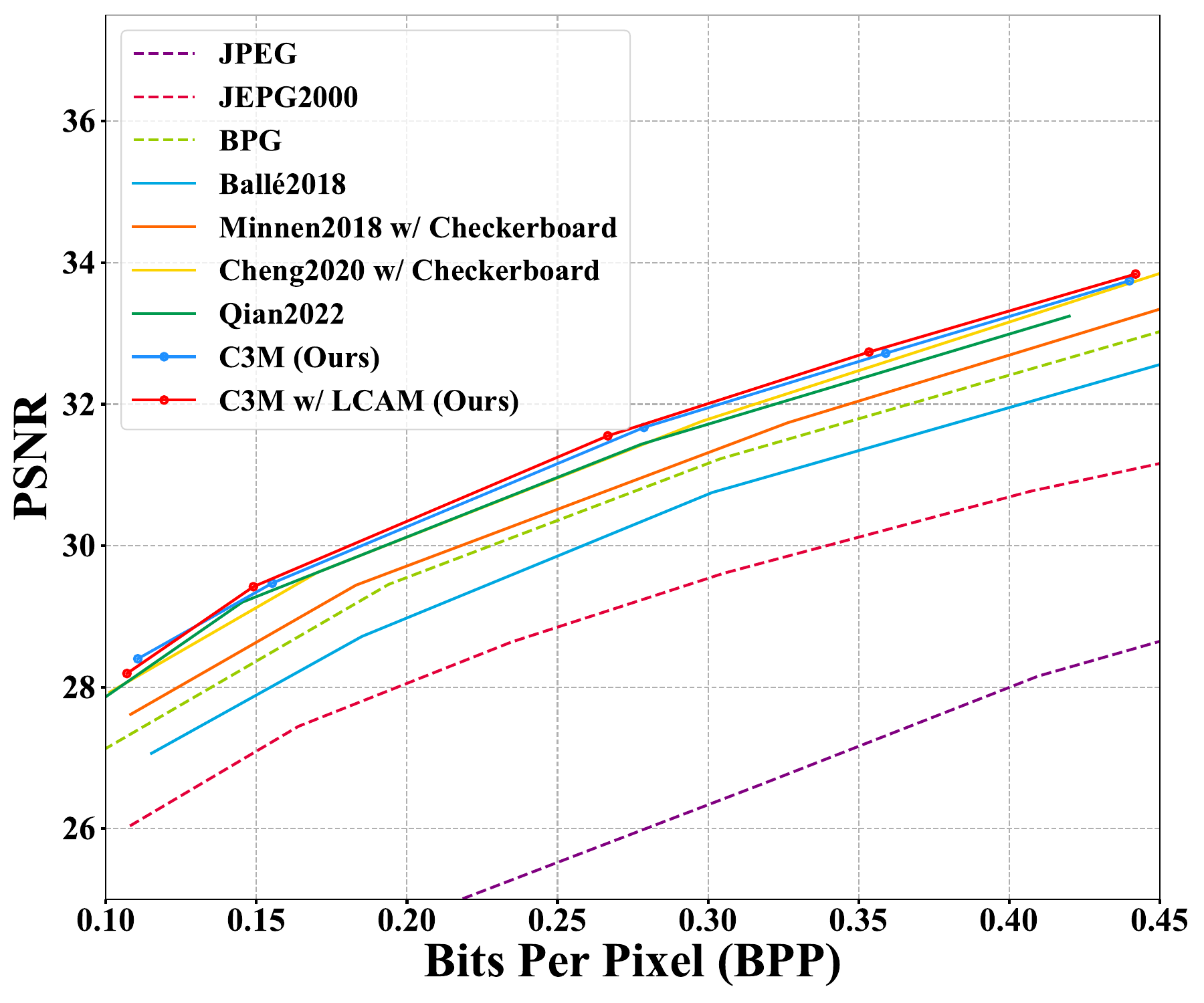}
    \caption{R-D performance averaged on Kodak dataset using MSE evaluated with proposed C$^3$M and LCAM.}
    \label{fig:kodak-mse-1}
\vspace{-2mm}
\end{figure}

\textbf{Evaluation.} We compare our proposed model with the traditional image compression algorithms (BPG and JPEG) as well as the learned image compression works \cite{minnen2018joint, cheng2020learned,he2021checkerboard,qian2021entroformer}. Different approaches are evaluated on the same Kodak dataset. In this study, we compare but do not emphasize the channel-wise context works \cite{he2022elic, minnen2020channel} since channel-wise context models are orthogonal and complementary directions to our research focus.

\textbf{Architecture.} For the main encoder/decoder, the depth of LCAMs in four sequential blocks are set to 2, 4, 6, and 2, respectively. For the hyper encoder/decoder, we use two sequential transformers in each LCAM module. Following the existing transformer works \cite{dosovitskiy2020image}, we set the expansion of 2 as the MLP layers. For the C$^3$M, the channel dimension is set to 384 with 6 heads, where each head has 64 channels in each Masked Self-Attention module. The depth in each C$^3$M is set to 6. The expansion factor is 2 in the MLP layers.

\begin{table}[t]
\setlength{\tabcolsep}{6pt}

\centering
\scalebox{0.7}{
\begin{tabular}{lcccc}
\toprule

\multirow{2}{*}{\makecell{Context \\ Model}} & \multirow{2}{*}{\makecell{Encoder \\ Decoder}} & \multirow{2}{*}{\makecell{Kodak \\ Dataset}} & \multirow{2}{*}{\makecell{Tenick \\ Dataset}} & \multirow{2}{*}{\makecell{CLIC \\ Dataset}}\\ & & & & \\

\midrule 
- & BPG & 0      & 0        & 0 \\ 
Checkerboard\cite{he2021checkerboard} & Minnen2018\cite{minnen2018joint}& -9.06      & -9.21        & -15.99 \\ 
Checkerboard\cite{he2021checkerboard} & Cheng2020\cite{cheng2020learned}  & -17.41      & -17.22    &  -22.91 \\
Checkerboard\cite{he2021checkerboard} & Qian2022\cite{qian2021entroformer} & -17.92      & -19.84    &  -23.56 \\

Checkerboard\cite{he2021checkerboard} & LCAM (Ours) & -18.96      & -20.91    &  -24.63 \\

AutoRegCNN\cite{minnen2018joint} & Minnen2018\cite{minnen2018joint} & -11.3      & -11.66        & -18.27 \\ 
AutoRegCNN\cite{minnen2018joint} & Cheng2020\cite{cheng2020learned}  & -17.73      & -17.69    &  -23.53 \\ 
AutoRegCNN\cite{minnen2018joint} & LCAM (Ours)  & -19.02      & -20.48    &  -24.79 \\ 

C$^3$M (Ours) & Minnen2018\cite{minnen2018joint} & -12.28      & -12.39      & -18.95 \\ 
C$^3$M (Ours) & Cheng2020\cite{cheng2020learned}  & -18.25      & -18.18    &  -24.32 \\
C$^3$M (Ours) & Qian2022\cite{qian2021entroformer} & -19.47      & -20.43    &  -25.97  \\

\textbf{C$^3$M (Ours)} & \textbf{LCAM (Ours)} & \textbf{-20.84}      & \textbf{-22.61}         & \textbf{-27.20} \\

\bottomrule
\end{tabular}
}
\caption{Averaged BD-rate (\%) improvement for different dataset.}
\label{table:ablation_c3m}
\vspace{-4mm}
\end{table}

\begin{table}[t]
\setlength{\tabcolsep}{4pt}
\centering
\scalebox{0.7}{
\begin{tabular}{lccccc}
\toprule

\multirow{2}{*}{Method} & \multirow{2}{*}{\makecell{Context \\ Type}} & \multirow{2}{*}{\makecell{Context \\ Model}} & \multirow{2}{*}{\makecell{Latency \\ (ms)}} & \multirow{2}{*}{\makecell{Time \\ Complexity}}\\ & & & & \\

\midrule 
Minnen2018\cite{minnen2018joint} & Parallel      & Checkerboard\cite{he2021checkerboard}                   & 47.6 & Two Pass\\
Cheng2020\cite{cheng2020learned}  & Parallel      & Checkerboard\cite{he2021checkerboard}                  & 68.3 & Two Pass\\ 
Minnen2018\cite{minnen2018joint} & Serial      & AutoRegCNN\cite{minnen2018joint}                   & $>$1000 & $\mathcal{O}(n^2)$\\
Cheng2020\cite{cheng2020learned}  & Serial      & AutoRegCNN\cite{minnen2018joint}                  & $>$1000 & $\mathcal{O}(n^2)$\\ 
Qian2022\cite{qian2021entroformer} & Serial      & AutoRegTrans\cite{qian2021entroformer}         & $>$1000 & $\mathcal{O}(n^2)$\\
Cheng2020\cite{cheng2020learned}  & Channel      & Channel\cite{minnen2020channel}                & 528.8 & - \\ 
Minnen2018\cite{minnen2018joint} & Parallel      & C$^3$M (Ours)                   & 242.5 & $\mathcal{O}(\log{}n)$\\
Cheng2020\cite{cheng2020learned}  & Parallel      & C$^3$M (Ours)                & 261.9 & $\mathcal{O}(\log{}n)$\\ 

\bottomrule
\end{tabular}
}
\caption{Deocoding latency(ms) on Kodak dataset.}
\label{table:decoding-time}
\vspace{-4mm}
\end{table}

\subsection{Results}
\label{sec:abl}

\textbf{Effect of C$^3$M and LCAM.} We conduct experiments to evaluate the effect of C$^3$M. As shown in Fig. \ref{fig:kodak-mse-1}, when using the same architecture that is adopted in Qian2022 \cite{qian2021entroformer}, the introduction of C$^3$M to context model can improve PSNR with the same bpp on the Kodak dataset. We also conduct the ablation studies by measuring the BD-rate improvement for different datasets: Kodak dataset, Tenick dataset, and CLIC dataset shown as Table \ref{table:ablation_c3m}. We employ Checkerboard\cite{he2021checkerboard} and AutoRegressiveCNN \cite{minnen2018joint} as the baseline context models and Minnen2018 \cite{minnen2018joint}, Cheng2020 \cite{cheng2020learned}, and Qian2022 \cite{qian2021entroformer} as the baseline models for Encoder/Decoder. Comparing our proposed C$^3$M and LCAM with these established baselines shows that both C$^3$M and LCAM exhibit superior performance individually. This demonstrates the effectiveness of our methods in capturing the critical features and dependencies necessary for improved image compression, highlighting the advantages of our approach over the existing techniques.

\textbf{Decoding Time.} 
We conduct experiments to evaluate the decoding time of the proposed C$^3$M on the Kodak dataset on the NVIDIA V100 GPUs. As shown in Table \ref{table:decoding-time}, compared to the Checkerboard, C$^3$M performs slower than it with $\mathcal{O}(\log{}n)$ time complexity. Nevertheless, it is essential to highlight that this is still a remarkable achievement. The proposed C$^3$M model offers a significant improvement in decoding time. Specifically, when decoding a 512 $\times$ 768 Kodak image, the C$^3$M model requires only 250ms, whereas the autoregressive context model takes more than 1000ms. This marked reduction in decoding time highlights the efficiency of the C$^3$M model in balancing computational demands and rate-distortion performance, making it a promising alternative to existing context models in the domain of learned image compression.

\begin{figure}[t]
    \centering
    \includegraphics[width=1\linewidth]{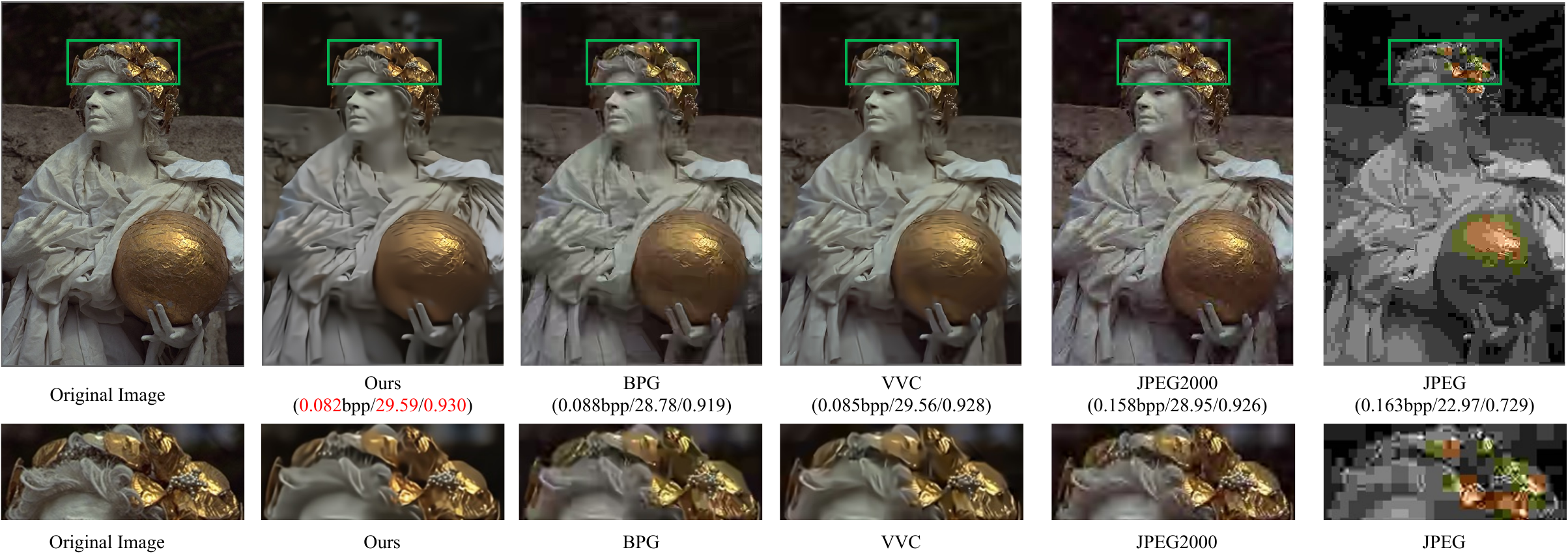}
    \caption{Visualization of reconstructed images and enlarged images of \textit{kodim17} from Kodak dataset.}
    \label{fig:vis1.}
    \vspace{-2mm}
\end{figure}

\begin{figure}[t]
    \centering
    \includegraphics[width=1\linewidth]{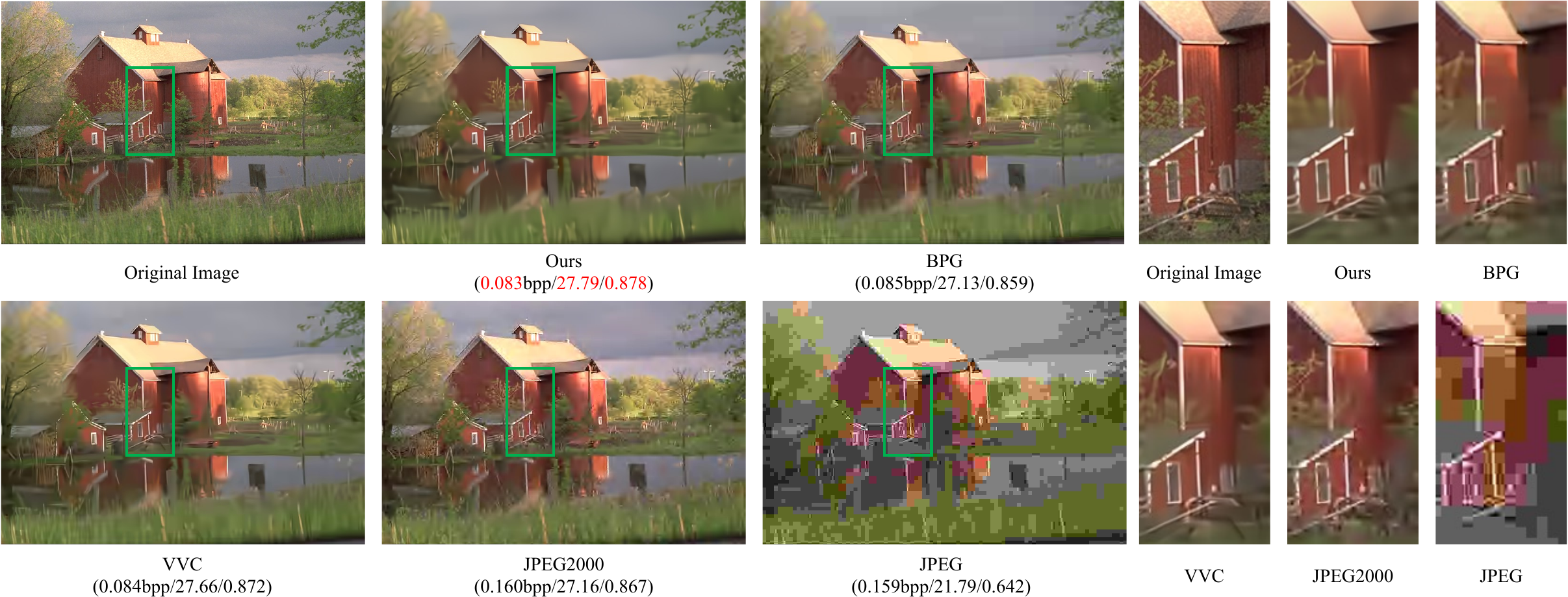}
    \caption{Visualization of reconstructed images and enlarged images of \textit{kodim22} from Kodak dataset.}
    \label{fig:vis2.}
\vspace{-2mm}
\end{figure}

\begin{figure}[t]
    \centering
        \includegraphics[width=0.7\linewidth]{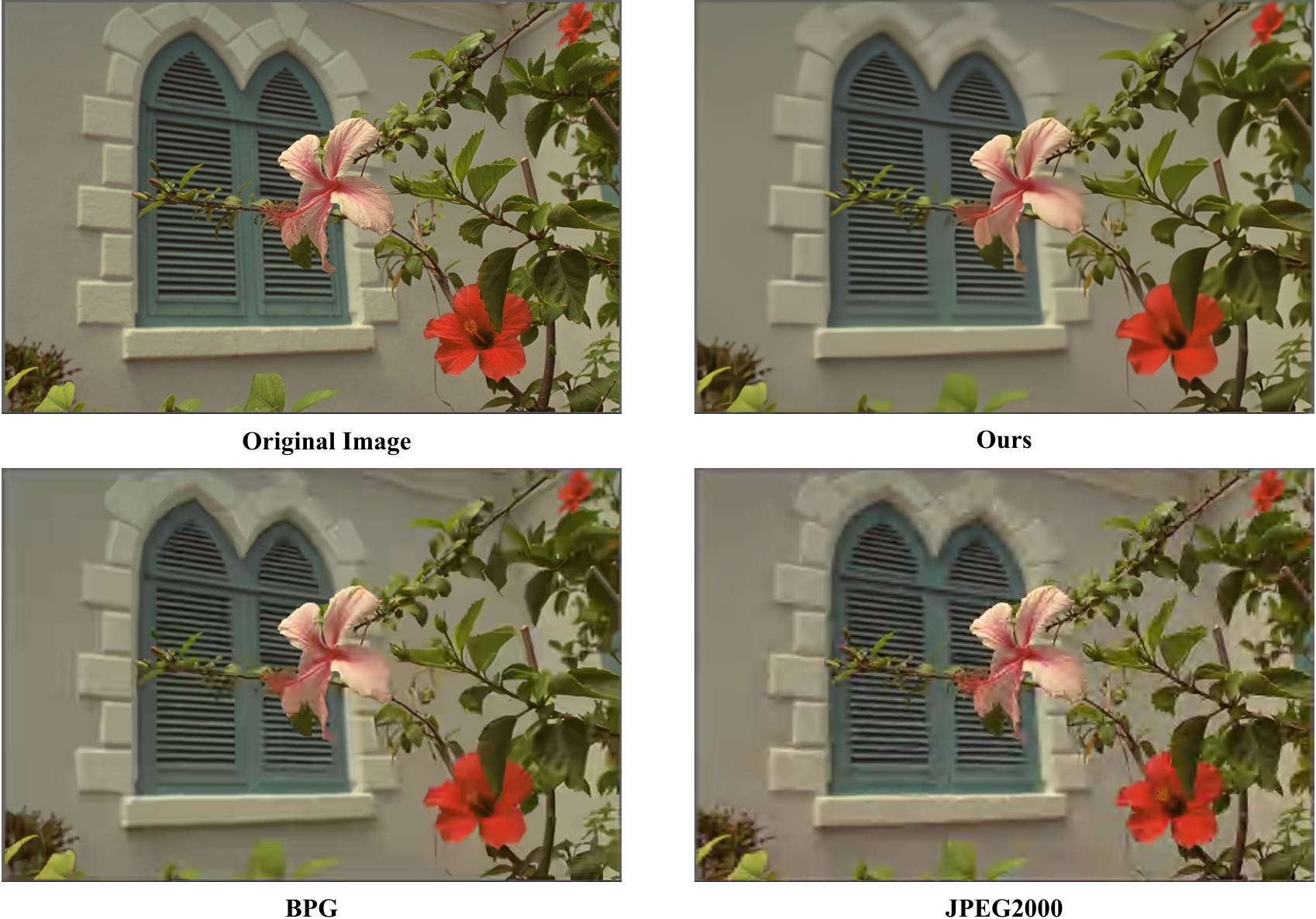} 
        \caption{Visualization of reconstructed images of \textit{kodim07} from Kodak dataset reveals notable differences upon zooming in. The image reconstructed using our method displays smoothly bricks alongside the window. In contrast, the reconstructions from traditional methods, such as JPEG2000, display distorted brick.}
        \label{fig:kodak07}
\vspace{-6mm}
\end{figure}

\begin{figure}[t]
    \centering
        \includegraphics[width=0.7\linewidth]{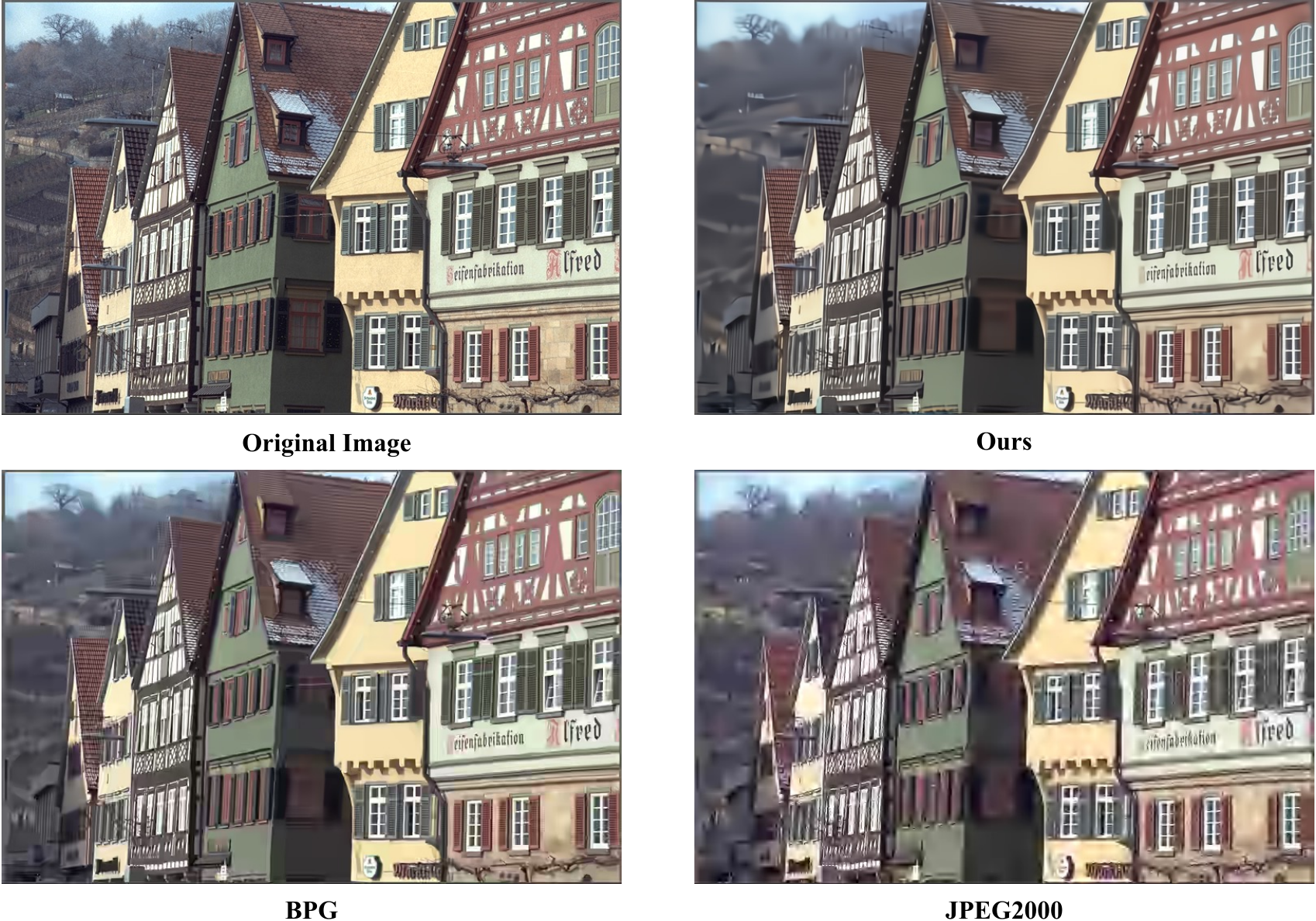}
        \caption{Visualization of reconstructed images of \textit{kodim08} from Kodak dataset reveals notable differences upon zooming in. The image reconstructed using our method displays a high-quality exterior wall. In contrast, those from traditional methods, such as JPEG2000, exhibit corrupted walls with obvious noise.}
        \label{fig:kodak08}
\vspace{-6mm}
\end{figure}

\textbf{Visualization.}
We visualize the reconstructed images from various approaches: JEPG, JPEG2000, BPG, VVC, and our MSE-optimized model with Fig. \ref{fig:vis1.}, Fig. \ref{fig:vis2.}, Fig. \ref{fig:kodak07}, and Fig. \ref{fig:kodak08}. Fig. \ref{fig:vis1.} shows the constructed image of KODIM17 with 0.088 bpp and 28.78 PSNR using the BPG method, 0.085 bpp and 29.56 using the VVC method, and 0.158 bpp and 28.95 PSNR using JPEG2000 method. It is seen that our approach can outperform the above coding methods by providing 0.082 bpp and 29.59 PSNR. Fig. \ref{fig:vis2.} shows the reconstructed image of KODIM22 with 0.085 bpp and 27.13 PSNR using the BPG method, 0.084 bpp and 27.66 using the VVC method, and 0.160 bpp and 27.16 PSNR using JPEG2000 method. Our model can outperform the above coding methods by providing 0.083 bpp and 27.79 PSNR. For the enlarged image in Fig. \ref{fig:vis2.}, our method reveals more details of texture information. For instance, the middle branch of the tree is missing when using the VVC method, while it is reserved using our approach. Also, compared with JPEG2000, our learning-based solution can reconstruct a smoother image with less noise. 

\textbf{Subjective Analysis.} From the visualization results, it is seen that our method still has some limitations. For the reconstructed images in Fig. \ref{fig:vis1.} and Fig. \ref{fig:vis2.}, our approach does not focus on drastic changes in detail. We hypothesize that this limitation may be attributed to the transformer-based backbone's inherent inclination towards capturing long-range dependencies and semantic information within images instead of concentrating on local texture information. More specifically, the self-attention mechanism utilized by transformers facilitates the consideration of global context when processing input images, which proves beneficial for identifying semantic relationships. In the context of image compression, the transformer-based backbone might prioritize retaining semantic information over preserving high-frequency details, such as local textures. This prioritization can improve rate-distortion performance, as the model effectively compresses the most semantically relevant information. Nonetheless, focusing on local texture information may result in losing details in the reconstructed images, potentially making them unsuitable for specific applications. We believe that enhancing the details of the reconstructed images by transformer-based backbone is an exciting research direction that can be explored in future works.

\section{Conclusion}
\label{sec:conclusion}

In this paper, we propose a corner-to-center context model for learned image compression. The motivation stems from the lack of causal context and limited receptive field used in the previous parallel context model. To address this limitation, we sequentially predict the context features in a logarithm-based corner-to-center order. Then, in order to further improve the capacity of capturing long-range dependencies and more information, we propose to enlarge the local window and adopt various window shapes in the encoder and decoder. Experiments show that our approach can outperform the state-of-the-art methods in terms of R-D performance. Our proposed context model only takes $\mathcal{O}(\log{}n)$ time complexity instead of $\mathcal{O}(n^2)$, achieving significant time savings in the decoding process compared to the serial context model, especially for the high-resolution image. We also perform subjective analysis and discuss the limitations of our proposed solution. We suggest that improving the detailed representation of reconstructed images is an interesting research direction for transformer-based image compression works in the future.

{\small
\bibliographystyle{ieee_fullname}
\bibliography{egbib}

\begin{thebibliography}{10}\itemsep=-1pt

\bibitem{agustsson2019generative}
Eirikur Agustsson, Michael Tschannen, Fabian Mentzer, Radu Timofte, and Luc~Van
  Gool.
\newblock Generative adversarial networks for extreme learned image
  compression.
\newblock In {\em Proceedings of the IEEE/CVF International Conference on
  Computer Vision}, pages 221--231, 2019.

\bibitem{balle2016end}
Johannes Ball{\'e}, Valero Laparra, and Eero~P Simoncelli.
\newblock End-to-end optimized image compression.
\newblock {\em arXiv preprint arXiv:1611.01704}, 2016.

\bibitem{balle2017endtoend}
Johannes Ball{\'e}, Valero Laparra, and Eero~P. Simoncelli.
\newblock End-to-end optimized image compression.
\newblock In {\em International Conference on Learning Representations}, 2017.

\bibitem{balle2018variational}
Johannes Ball{\'e}, David Minnen, Saurabh Singh, Sung~Jin Hwang, and Nick
  Johnston.
\newblock Variational image compression with a scale hyperprior.
\newblock In {\em International Conference on Learning Representations}, 2018.

\bibitem{begaint2020compressai}
Jean B{\'e}gaint, Fabien Racap{\'e}, Simon Feltman, and Akshay Pushparaja.
\newblock Compressai: a pytorch library and evaluation platform for end-to-end
  compression research.
\newblock {\em arXiv preprint arXiv:2011.03029}, 2020.

\bibitem{bellard2016bpg}
Fabrice Bellard.
\newblock Bpg image format (2014).
\newblock {\em Volume}, 1:2, 2016.

\bibitem{bross2021overview}
Benjamin Bross, Ye-Kui Wang, Yan Ye, Shan Liu, Jianle Chen, Gary~J Sullivan,
  and Jens-Rainer Ohm.
\newblock Overview of the versatile video coding (vvc) standard and its
  applications.
\newblock {\em IEEE Transactions on Circuits and Systems for Video Technology},
  31(10):3736--3764, 2021.

\bibitem{cheng2020learned}
Zhengxue Cheng, Heming Sun, Masaru Takeuchi, and Jiro Katto.
\newblock Learned image compression with discretized gaussian mixture
  likelihoods and attention modules.
\newblock In {\em Proceedings of the IEEE/CVF Conference on Computer Vision and
  Pattern Recognition}, pages 7939--7948, 2020.

\bibitem{choi2019variable}
Yoojin Choi, Mostafa El-Khamy, and Jungwon Lee.
\newblock Variable rate deep image compression with a conditional autoencoder.
\newblock In {\em Proceedings of the IEEE/CVF International Conference on
  Computer Vision}, pages 3146--3154, 2019.

\bibitem{cui2021asymmetric}
Ze Cui, Jing Wang, Shangyin Gao, Tiansheng Guo, Yihui Feng, and Bo Bai.
\newblock Asymmetric gained deep image compression with continuous rate
  adaptation.
\newblock In {\em Proceedings of the IEEE/CVF Conference on Computer Vision and
  Pattern Recognition}, pages 10532--10541, 2021.

\bibitem{dosovitskiy2020image}
Alexey Dosovitskiy, Lucas Beyer, Alexander Kolesnikov, Dirk Weissenborn,
  Xiaohua Zhai, Thomas Unterthiner, Mostafa Dehghani, Matthias Minderer, Georg
  Heigold, Sylvain Gelly, et~al.
\newblock An image is worth 16x16 words: Transformers for image recognition at
  scale.
\newblock In {\em International Conference on Learning Representations}, 2020.

\bibitem{dosovitskiy2019you}
Alexey Dosovitskiy and Josip Djolonga.
\newblock You only train once: Loss-conditional training of deep networks.
\newblock In {\em International conference on learning representations}, 2019.

\bibitem{dumoulin2016guide}
Vincent Dumoulin and Francesco Visin.
\newblock A guide to convolution arithmetic for deep learning.
\newblock {\em arXiv preprint arXiv:1603.07285}, 2016.

\bibitem{d2021convit}
St{\'e}phane d’Ascoli, Hugo Touvron, Matthew~L Leavitt, Ari~S Morcos, Giulio
  Biroli, and Levent Sagun.
\newblock Convit: Improving vision transformers with soft convolutional
  inductive biases.
\newblock In {\em International Conference on Machine Learning}, pages
  2286--2296. PMLR, 2021.

\bibitem{golts2021image}
Alex Golts and Yoav~Y Schechner.
\newblock Image compression optimized for 3d reconstruction by utilizing deep
  neural networks.
\newblock {\em Journal of Visual Communication and Image Representation},
  79:103208, 2021.

\bibitem{graham2021levit}
Benjamin Graham, Alaaeldin El-Nouby, Hugo Touvron, Pierre Stock, Armand Joulin,
  Herv{\'e} J{\'e}gou, and Matthijs Douze.
\newblock Levit: a vision transformer in convnet's clothing for faster
  inference.
\newblock In {\em Proceedings of the IEEE/CVF international conference on
  computer vision}, pages 12259--12269, 2021.

\bibitem{he2022elic}
Dailan He, Ziming Yang, Weikun Peng, Rui Ma, Hongwei Qin, and Yan Wang.
\newblock Elic: Efficient learned image compression with unevenly grouped
  space-channel contextual adaptive coding.
\newblock In {\em Proceedings of the IEEE/CVF Conference on Computer Vision and
  Pattern Recognition}, pages 5718--5727, 2022.

\bibitem{he2021checkerboard}
Dailan He, Yaoyan Zheng, Baocheng Sun, Yan Wang, and Hongwei Qin.
\newblock Checkerboard context model for efficient learned image compression.
\newblock In {\em Proceedings of the IEEE/CVF Conference on Computer Vision and
  Pattern Recognition}, pages 14771--14780, 2021.

\bibitem{hinton2006reducing}
Geoffrey~E Hinton and Ruslan~R Salakhutdinov.
\newblock Reducing the dimensionality of data with neural networks.
\newblock {\em science}, 313(5786):504--507, 2006.

\bibitem{johnston2018improved}
Nick Johnston, Damien Vincent, David Minnen, Michele Covell, Saurabh Singh,
  Troy Chinen, Sung~Jin Hwang, Joel Shor, and George Toderici.
\newblock Improved lossy image compression with priming and spatially adaptive
  bit rates for recurrent networks.
\newblock In {\em Proceedings of the IEEE Conference on Computer Vision and
  Pattern Recognition}, pages 4385--4393, 2018.

\bibitem{kim2022joint}
Jun-Hyuk Kim, Byeongho Heo, and Jong-Seok Lee.
\newblock Joint global and local hierarchical priors for learned image
  compression.
\newblock In {\em Proceedings of the IEEE/CVF Conference on Computer Vision and
  Pattern Recognition}, pages 5992--6001, 2022.

\bibitem{koyuncu2022contextformer}
A~Burakhan Koyuncu, Han Gao, and Eckehard Steinbach.
\newblock Contextformer: A transformer with spatio-channel attention for
  context modeling in learned image compression.
\newblock {\em arXiv preprint arXiv:2203.02452}, 2022.

\bibitem{lee2018context}
Jooyoung Lee, Seunghyun Cho, and Seung-Kwon Beack.
\newblock Context-adaptive entropy model for end-to-end optimized image
  compression.
\newblock In {\em International Conference on Learning Representations}, 2018.

\bibitem{liu2019non}
Haojie Liu, Tong Chen, Peiyao Guo, Qiu Shen, Xun Cao, Yao Wang, and Zhan Ma.
\newblock Non-local attention optimized deep image compression.
\newblock {\em arXiv preprint arXiv:1904.09757}, 2019.

\bibitem{liu2020unified}
Jiaheng Liu, Guo Lu, Zhihao Hu, and Dong Xu.
\newblock A unified end-to-end framework for efficient deep image compression.
\newblock {\em arXiv preprint arXiv:2002.03370}, 2020.

\bibitem{marcellin2000overview}
Michael~W Marcellin, Michael~J Gormish, Ali Bilgin, and Martin~P Boliek.
\newblock An overview of jpeg-2000.
\newblock In {\em Proceedings DCC 2000. Data Compression Conference}, pages
  523--541. IEEE, 2000.

\bibitem{mentzer2018conditional}
Fabian Mentzer, Eirikur Agustsson, Michael Tschannen, Radu Timofte, and Luc
  Van~Gool.
\newblock Conditional probability models for deep image compression.
\newblock In {\em Proceedings of the IEEE Conference on Computer Vision and
  Pattern Recognition}, pages 4394--4402, 2018.

\bibitem{minnen2018joint}
David Minnen, Johannes Ball{\'e}, and George~D Toderici.
\newblock Joint autoregressive and hierarchical priors for learned image
  compression.
\newblock {\em Advances in neural information processing systems}, 31, 2018.

\bibitem{minnen2020channel}
David Minnen and Saurabh Singh.
\newblock Channel-wise autoregressive entropy models for learned image
  compression.
\newblock In {\em 2020 IEEE International Conference on Image Processing
  (ICIP)}, pages 3339--3343. IEEE, 2020.

\bibitem{naseer2021intriguing}
Muhammad~Muzammal Naseer, Kanchana Ranasinghe, Salman~H Khan, Munawar Hayat,
  Fahad Shahbaz~Khan, and Ming-Hsuan Yang.
\newblock Intriguing properties of vision transformers.
\newblock {\em Advances in Neural Information Processing Systems},
  34:23296--23308, 2021.

\bibitem{qian2021entroformer}
Yichen Qian, Xiuyu Sun, Ming Lin, Zhiyu Tan, and Rong Jin.
\newblock Entroformer: A transformer-based entropy model for learned image
  compression.
\newblock In {\em International Conference on Learning Representations}, 2021.

\bibitem{qian2020learning}
Yichen Qian, Zhiyu Tan, Xiuyu Sun, Ming Lin, Dongyang Li, Zhenhong Sun, Li Hao,
  and Rong Jin.
\newblock Learning accurate entropy model with global reference for image
  compression.
\newblock In {\em International Conference on Learning Representations}, 2020.

\bibitem{raghu2021vision}
Maithra Raghu, Thomas Unterthiner, Simon Kornblith, Chiyuan Zhang, and Alexey
  Dosovitskiy.
\newblock Do vision transformers see like convolutional neural networks?
\newblock {\em Advances in Neural Information Processing Systems},
  34:12116--12128, 2021.

\bibitem{sun2021interpolation}
Zhenhong Sun, Zhiyu Tan, Xiuyu Sun, Fangyi Zhang, Yichen Qian, Dongyang Li, and
  Hao Li.
\newblock Interpolation variable rate image compression.
\newblock In {\em Proceedings of the 29th ACM International Conference on
  Multimedia}, pages 5574--5582, 2021.

\bibitem{taubman2002jpeg2000}
David~S Taubman and Michael~W Marcellin.
\newblock Jpeg2000: Image compression fundamentals.
\newblock {\em Standards and Practice}, 11(2), 2002.

\bibitem{theis2017lossy}
Lucas Theis, Wenzhe Shi, Andrew Cunningham, and Ferenc Husz{\'a}r.
\newblock Lossy image compression with compressive autoencoders.
\newblock In {\em International Conference on Learning Representations}, 2017.

\bibitem{toderici2017full}
George Toderici, Damien Vincent, Nick Johnston, Sung Jin~Hwang, David Minnen,
  Joel Shor, and Michele Covell.
\newblock Full resolution image compression with recurrent neural networks.
\newblock In {\em Proceedings of the IEEE conference on Computer Vision and
  Pattern Recognition}, pages 5306--5314, 2017.

\bibitem{van2016conditional}
Aaron Van~den Oord, Nal Kalchbrenner, Lasse Espeholt, Oriol Vinyals, Alex
  Graves, et~al.
\newblock Conditional image generation with pixelcnn decoders.
\newblock {\em Advances in neural information processing systems}, 29, 2016.

\bibitem{wallace1991jpeg}
Gregory~K Wallace.
\newblock The jpeg still picture compression standard.
\newblock {\em Communications of the ACM}, 34(4):30--44, 1991.

\bibitem{xiao2021early}
Tete Xiao, Mannat Singh, Eric Mintun, Trevor Darrell, Piotr Doll{\'a}r, and
  Ross Girshick.
\newblock Early convolutions help transformers see better.
\newblock {\em Advances in Neural Information Processing Systems},
  34:30392--30400, 2021.

\bibitem{zou2022devil}
Renjie Zou, Chunfeng Song, and Zhaoxiang Zhang.
\newblock The devil is in the details: Window-based attention for image
  compression.
\newblock In {\em Proceedings of the IEEE/CVF Conference on Computer Vision and
  Pattern Recognition}, pages 17492--17501, 2022.

\end{thebibliography}
}

\clearpage
\section*{Appendix}

\section*{Details of Model}

\textbf{Decoder Pipeline.} 
To recover the constructed image $\bh{x} \in \mathbb{R}^{H\times{W}\times{C}}$ from decoded latents $\bs{y} \in \mathbb{R}^{\frac{H}{16}\times\frac{W}{16}}\times{C}$, we use the reverse architecture of the encoder. We first conduct several LCAMs with the input $\bh{y} \in \mathbb{R}^{\frac{H}{16}\times\frac{W}{16}}\times{C}$ and get the immediate output $\bs{o} \in \mathbb{R}^{\frac{H}{16}\times\frac{W}{16}}\times{C}$. Then, we send output $\bs{o}$ to the transposed convolutional layer to upscale the image $\bh{y} \in \mathbb{R}^{\frac{H}{8}\times\frac{W}{8}}\times{C}$. Here, the transposed convolution layer \cite{dumoulin2016guide} acts as an upscale layer, as opposed to the downsample convolutional layer. Transposed convolution layers are set to 3$\times$3 kernel size, 128 channels with the stride of 2. Then, repeatedly, the $\bs{y}_1$ is fed into the several same LCAMs with a convolution layer with the mirror configuration in the encoder. The channels decrease from 192 to 128. Finally, we can get the reconstructed image $\bh{x} \in \mathbb{R}^{{H}\times{W}\times{C}}$.

\textbf{Hyper Encoder/Decoder Pipeline.} 
Similar to the main encoder, we downsample the latents $\bh{y}$ with the downsample convolutional layer and LCAMs module (shown as Fig. \ref{fig:lcam}) in the hyper encoder. And we upscale the latents $\bh{z}$ with the transposed upscale convolutional layer and LCAMs module in the hyper decoder. The depth of LCAMs is set to 2 and 2 in two repeated blocks, respectively, for both hyper encoder/decoder. 

\section*{Additional Results}
\begin{figure}[ht]
    \centering
    \includegraphics[width=0.7\linewidth]{ 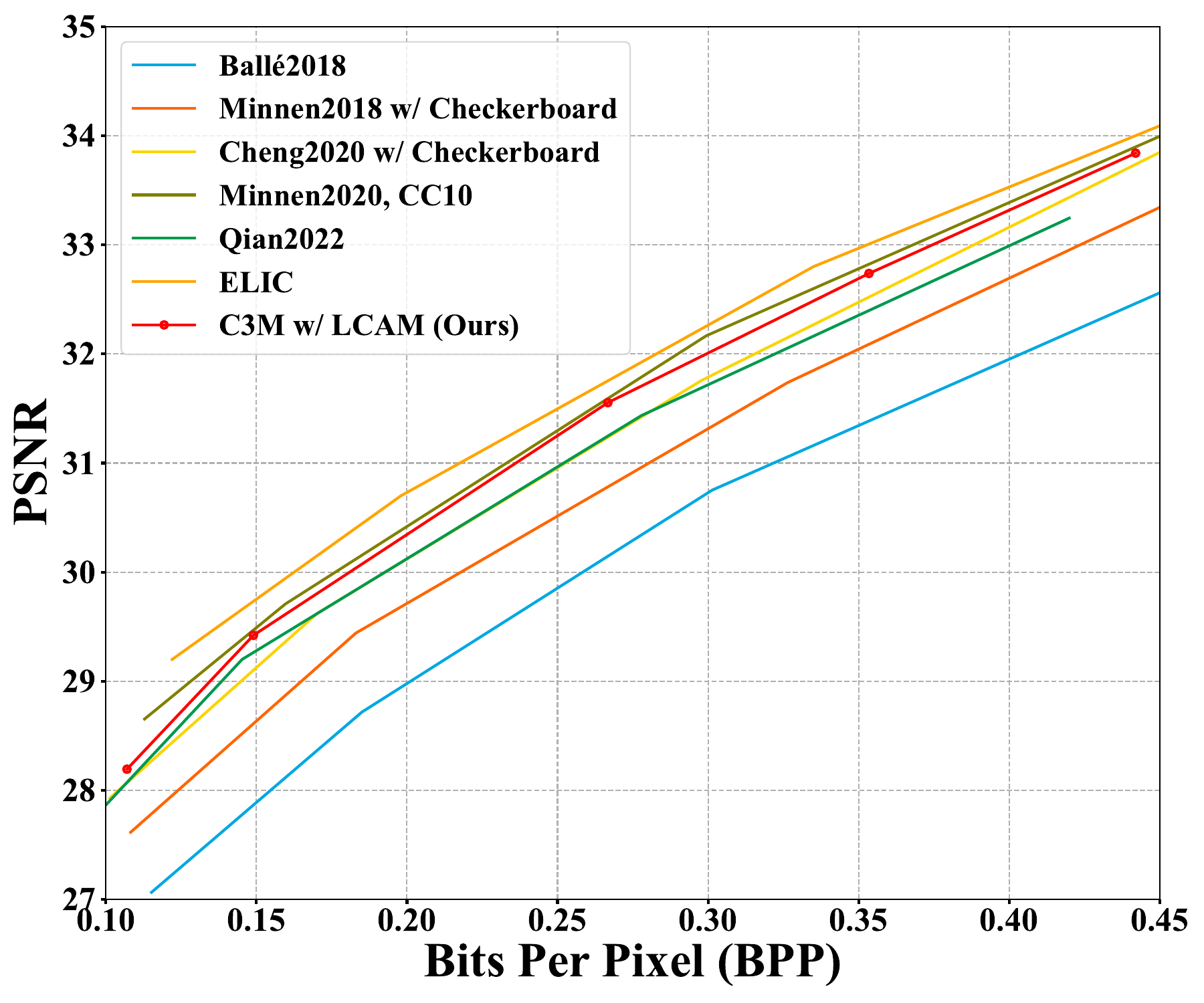}
    \caption{Comparisons between the channel-wise LIC methods Minnen2020-CC10 \cite{minnen2020channel}, ELIC \cite{he2022elic} and ours. R-D performance averaged on Kodak dataset using MSE loss.}
    \label{fig:channel.}
\end{figure}

\textbf{Comparisons with the channel-wise LIC methods.} We also show the results compared to the works \cite{minnen2020channel, he2022elic} with channel-wise context model. As depicted in Fig. \ref{fig:channel.}, channel-wise works \cite{minnen2020channel, he2022elic} outperforms ours, largely attributed to the additional decoding phase in the channel dimension. Incorporating channel-wise context prediction into our method offers an intriguing direction for future enhancements.

\section*{List of Symbols}
\begin{table}[ht]
\setlength{\tabcolsep}{1pt}
\centering
\scalebox{1}{
\begin{tabular}{cc}
Component                       & Symbol   \\
\toprule 
Input Image                     & $\boldsymbol{x}$              \\
Encoder                         & $g_a(\cdot)$                    \\
Latents                         & $\boldsymbol{y} = g_a(\boldsymbol{x})$ \\
Quantized Latents               & $\boldsymbol{\hat{y}} = Q(\boldsymbol{y})$       \\
Decoder                         & $g_s(\cdot)$                       \\
Hyper Encoder                   & $h_a(\cdot)$               \\
Hyper-latents                   & $\boldsymbol{z} = h_a(\boldsymbol{y})$                \\             
Quantized Hyper-latents         & $\boldsymbol{\hat{z}} = Q(\boldsymbol{z})$         \\
Hyper Decoder                   & $h_s(\cdot)$          \\
Context Parameters              & $\boldsymbol{\psi}=h_s(\bh{z})$    \\
Causal Context                  & $\bh{y}_{<i}$          \\
Context Features                & $\boldsymbol{\phi}=g_{cm}(\bh{y}_{<{i}}, \boldsymbol{\psi})$              \\
Entropy Parameter Network       & $g_{ep}(\cdot)$               \\
Mean, Scale                 & $\boldsymbol{\mu},\boldsymbol{\sigma}=g_{ep}(\boldsymbol{\psi}, \boldsymbol{\phi})$  \\
Reconstruction Image            & $\boldsymbol{\hat{x}}=g_s(\bh{y})$      \\      
\end{tabular}
}
\caption{List of Symbols}
\vspace{-4mm}
\end{table}

\section*{Enlarged and additional Visualization Results}

\begin{figure*}[t]
    \centering
    \includegraphics[width=1\linewidth]{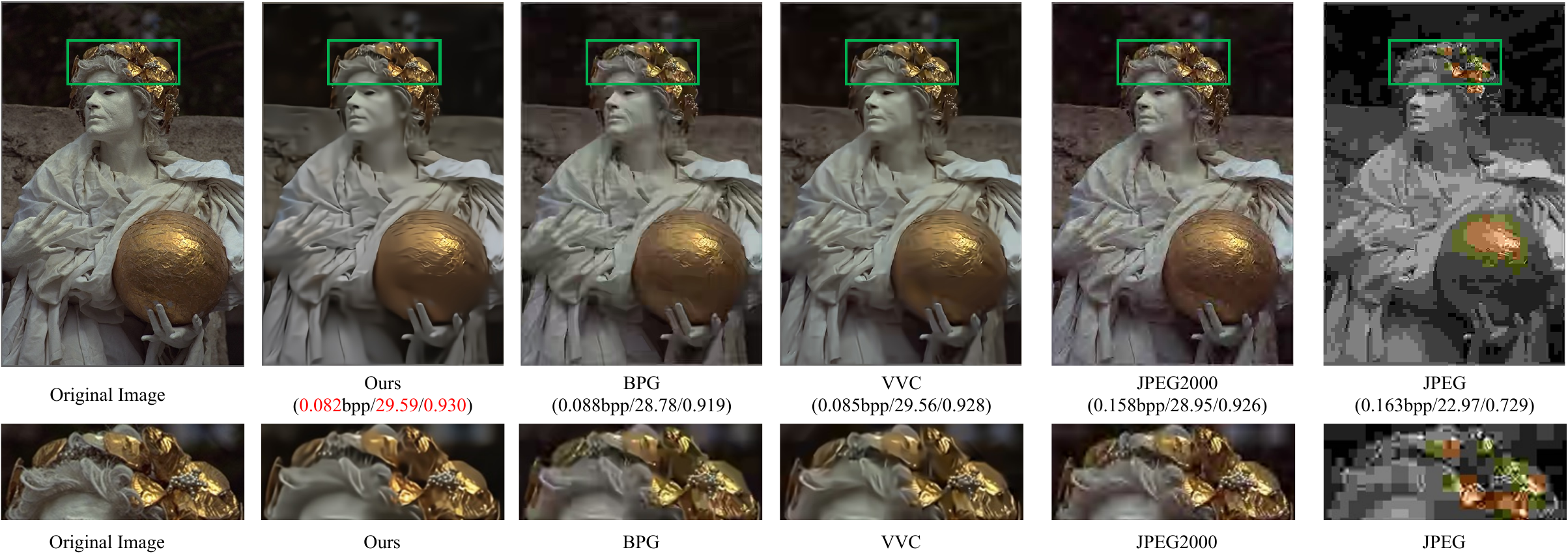}
    \caption{Visualization of reconstructed images and enlarged images of \textit{kodim17} from Kodak dataset.}
    \label{fig:vis1_1.}
    \vspace{-4mm}
\end{figure*}

\begin{figure*}[t]
    \centering
    \includegraphics[width=1\linewidth]{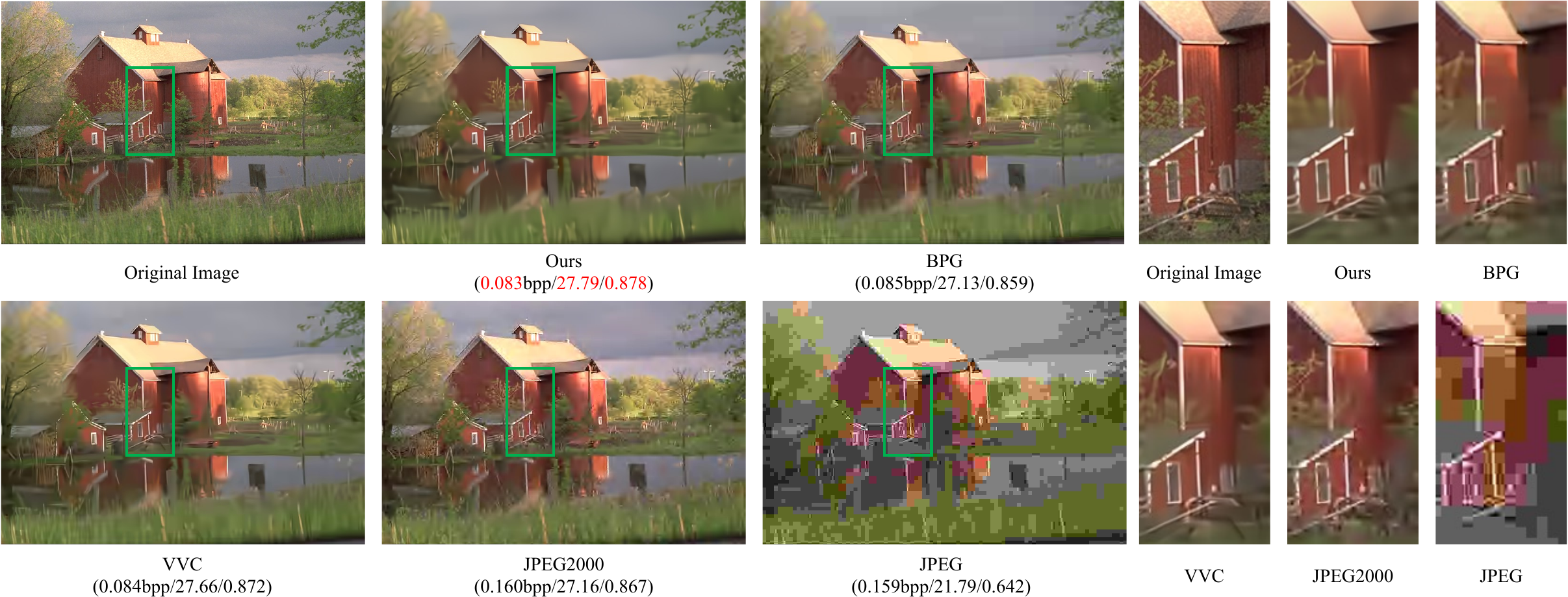}
    \caption{Visualization of reconstructed images and enlarged images of \textit{kodim22} from Kodak dataset.}
    \label{fig:vis2_1.}
\vspace{-4mm}
\end{figure*}

\begin{figure*}[t]
    \centering
        \includegraphics[width=0.9\linewidth]{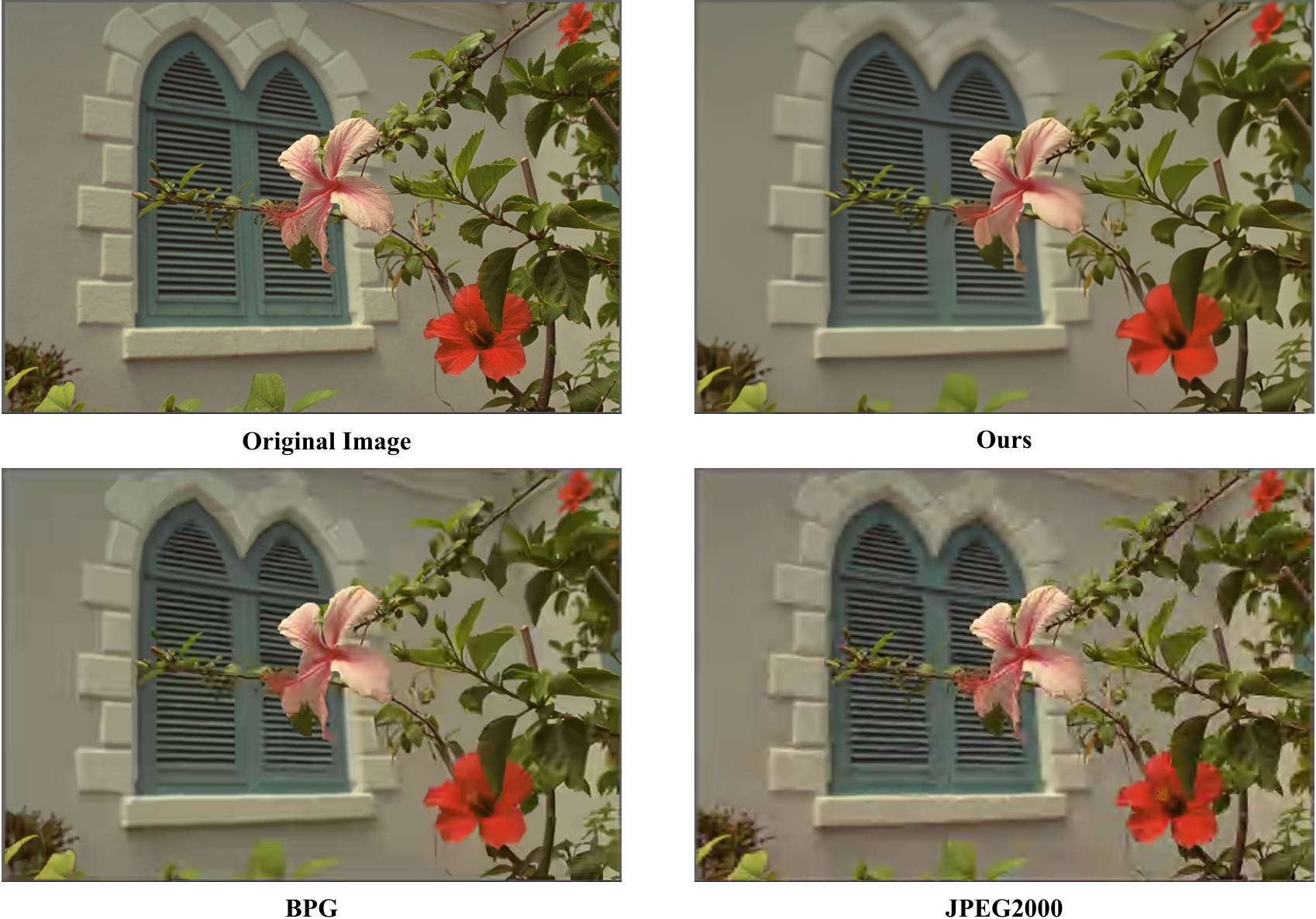} 
        \caption{Visualization of reconstructed images of \textit{kodim07} from Kodak dataset reveals notable differences upon zooming in. The image reconstructed using our method displays smoothly shaped bricks alongside the window. In contrast, the reconstructions from traditional methods, such as JPEG2000, display distorted brick shapes.}
        \label{fig:kodak07_1}
\vspace{-4mm}
\end{figure*}

\begin{figure*}[t]
    \centering
        \includegraphics[width=0.9\linewidth]{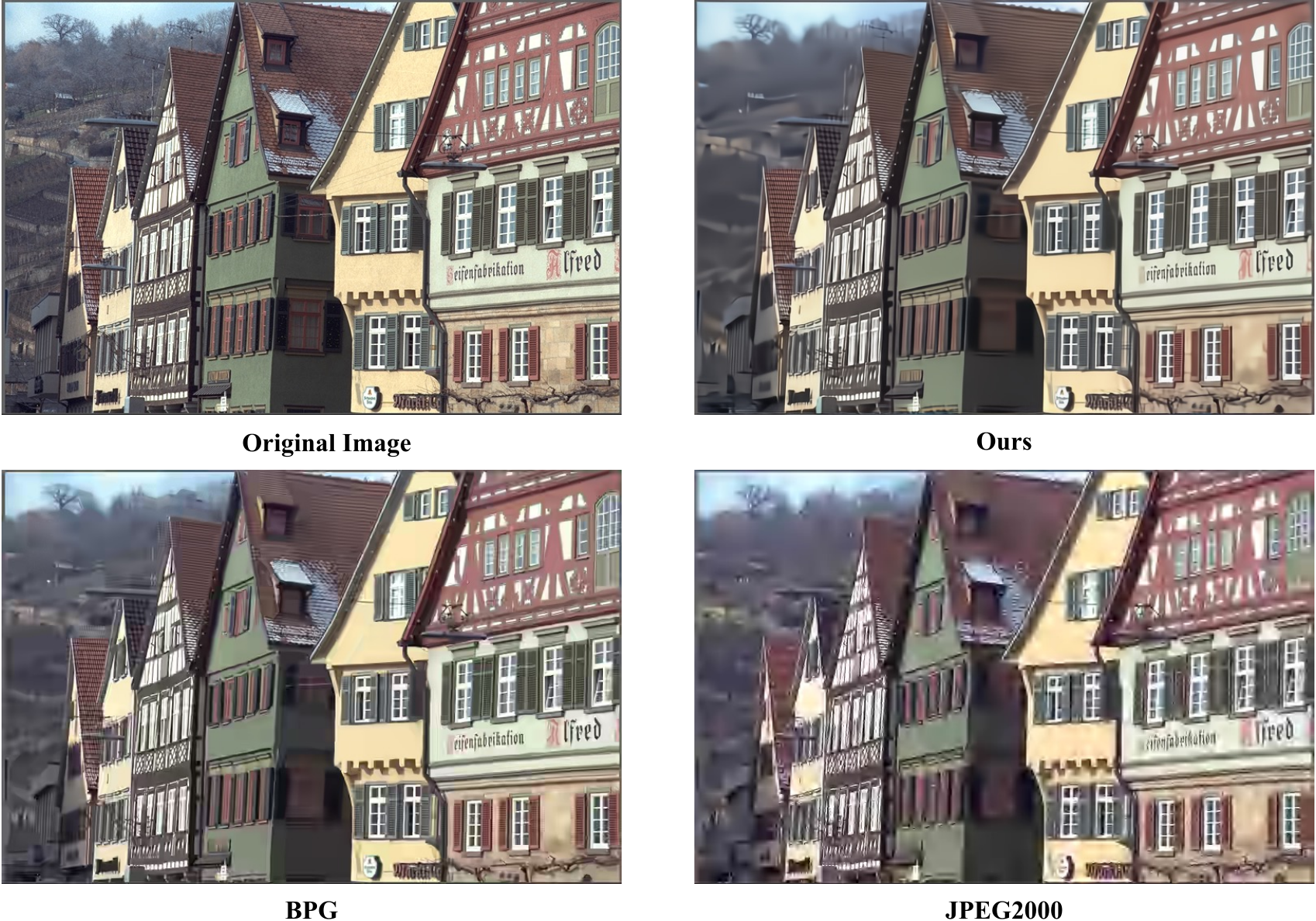}
        \caption{Visualization of reconstructed images of \textit{kodim08} from Kodak dataset reveals notable differences upon zooming in. The image reconstructed using our method displays a high-quality exterior wall. In contrast, the reconstructions from traditional methods, such as JPEG2000, exhibit corrupted walls with obvious noise.}
        \label{fig:kodak08_1}
\vspace{-4mm}
\end{figure*}


\end{document}


\title{Appendix}

\author{Yang Sui, Ding Ding, Xiang Pan, Xiaozhong Xu, Shan Liu, Bo Yuan, Zhenzhong Chen}

\maketitle

\section{Details of C$^3$M Backbone}

For the details of $\text{C}^3$M, we leverage the Masked Transformer model as our backbone. Given the reshaped input $\boldsymbol{\hat{y}} \in \mathbb{R}^{N \times d}$, where $N=H\times{W}$ denotes the total number of tokens sent into C$^3$M. Suppose the Masked Transformer model with the $\text{depth} = 1$, the context features $\boldsymbol{\phi}$ are predicted following:
\begin{equation}
\begin{aligned}
\boldsymbol{x^l} &= \text{Masked-MSA}(\text{LN}(\bh{y})) + \bh{y} \\
\boldsymbol{\phi} &= \text{MLP}(\text{LN}(\boldsymbol{x^l})) + \boldsymbol{x^l}
\end{aligned}
\label{eqn:transformer}
\end{equation}
where the attention module in Masked-MSA is following \cite{qian2021entroformer}:
\begin{equation}
\begin{aligned}
\text{Attn}(\boldsymbol{Q},\boldsymbol{K},\boldsymbol{V}) = \text{softmax}(\frac{\boldsymbol{Q}\boldsymbol{K}^{\text{T}} + \boldsymbol{P}}{\sqrt{d}})\boldsymbol{V},
\end{aligned}
\label{eqn:trans}
\end{equation}
where $\boldsymbol{P}$ denotes the position encoding and we only calculate the $\boldsymbol{\phi}$. In our experiments, we stack the Masked Transformer model with $\text{depth} = 6$. It is worth noting that in the last stage of prediction, we adopt the 3$\times$3 CNN as the backbone context model with better capturing the nearby local texture information motivated by nature between CNNs and Transformer \cite{xiao2021early}.

\begin{figure}[h]
    \centering
    \includegraphics[width=1\linewidth]{latex/fig/crossing-attention.pdf}
    \caption{Long-range Crossing Attention Module.}
    \label{fig:applcam}
\end{figure}

\section{Decoder, Hyper Encoder, Hyper Decoder}

\textbf{Decoder Pipeline.} 
To recover the constructed image $\bh{x} \in \mathbb{R}^{H\times{W}\times{C}}$ from decoded latents $\bs{y} \in \mathbb{R}^{\frac{H}{16}\times\frac{W}{16}}\times{C}$, we use the reverse architecture of the encoder. We first conduct several LCAMs with the input $\bh{y} \in \mathbb{R}^{\frac{H}{16}\times\frac{W}{16}}\times{C}$ and get the immediate output $\bs{o} \in \mathbb{R}^{\frac{H}{16}\times\frac{W}{16}}\times{C}$. Then, we send output $\bs{o}$ to the transposed convolutional layer to upscale the image $\bh{y} \in \mathbb{R}^{\frac{H}{8}\times\frac{W}{8}}\times{C}$. Here, the transposed convolution layer \cite{dumoulin2016guide} acts as an upscale layer, as opposed to the downsample convolutional layer. Transposed convolution layers are set to 3$\times$3 kernel size, 128 channels with the stride of 2. Then, repeatedly, the $\bs{y}_1$ is fed into the several same several LCAMs with a convolution layer with the mirror configuration in the encoder. The channels decrease from 192 to 128. Finally, we can get the reconstructed image $\bh{x} \in \mathbb{R}^{{H}\times{W}\times{C}}$.

\textbf{Hyper Encoder/Decoder Pipeline.} 
Similar to the main encoder, we downsample the latents $\bh{y}$ with the downsample convolutional layer and LCAMs module (shown as Fig. \ref{fig:applcam}) in the hyper encoder. And we upscale the latents $\bh{z}$ with the transposed upscale convolutional layer and LCAMs module in the hyper decoder. The depth of LCAMs is set to 2 and 2 in two repeated blocks, respectively, for both hyper encoder/decoder. 

\textbf{Visualization.}
We visualize the reconstructed images from various approaches: JPEG2000, BPG, and our proposed method. Fig. \ref{fig:kodak07}, \ref{fig:kodak08}, \ref{fig:kodak20} present the comparisons on reconstructed images of \textit{kodim07}, \textit{kodim08}, \textit{kodim20} from Kodak dataset, respectively.

\begin{figure*}[t]
    \centering
    \includegraphics[width=1\linewidth]{latex/fig/vis3.pdf}
    \caption{Visualization of reconstructed images of \textit{kodim07} from Kodak dataset.}
    \label{fig:kodak07}
\end{figure*}

\begin{figure*}[t]
    \centering
    \includegraphics[width=1\linewidth]{latex/fig/vis4.pdf}
    \caption{Visualization of reconstructed images of \textit{kodim08} from Kodak dataset.}
    \label{fig:kodak08}
\end{figure*}

\begin{figure*}[t]
    \centering
    \includegraphics[width=1\linewidth]{latex/fig/vis5.pdf}
    \caption{Visualization of reconstructed images of \textit{kodim20} from Kodak dataset.}
    \label{fig:kodak20}
\end{figure*}

\clearpage
{\small
\bibliographystyle{ieee_fullname}
\bibliography{egbib}
}